\documentclass[modern]{aastex63}

\usepackage{amsmath}
\usepackage{amssymb}
\usepackage{rotating}

\newcommand\tE{t_{\rm E}}

\newcommand{\ra}[4]{${#1}^{\rm h}{#2}^{\rm m}{#3}\fs{#4}$}
\newcommand{\dec}[4]{${#1}\arcdeg{#2}\arcmin{#3}\farcs{#4}$}

\newcommand\Fs{F_{\rm s}}
\newcommand\Fb{F_{\rm b}}

\received{January 1, 2018}
\revised{January 1, 2018}
\accepted{January 1, 2018}

\shorttitle{Microlensing in the Galactic plane}
\shortauthors{Mr\'oz et al.}

\begin{document}

\title{Microlensing optical depth and event rate in the OGLE-IV Galactic plane fields}

\correspondingauthor{Przemek Mr\'oz}
\email{pmroz@astro.caltech.edu}

\author[0000-0001-7016-1692]{Przemek Mr\'oz}
\affil{Division of Physics, Mathematics, and Astronomy, California Institute of Technology, Pasadena, CA 91125, USA}
\affil{Astronomical Observatory, University of Warsaw, Al. Ujazdowskie 4, 00-478 Warszawa, Poland}

\author[0000-0001-5207-5619]{Andrzej Udalski}
\affil{Astronomical Observatory, University of Warsaw, Al. Ujazdowskie 4, 00-478 Warszawa, Poland}

\author[0000-0002-0548-8995]{Micha\l{} K. Szyma\'nski}
\affil{Astronomical Observatory, University of Warsaw, Al. Ujazdowskie 4, 00-478 Warszawa, Poland}

\author[0000-0002-7777-0842]{Igor Soszy\'nski}
\affil{Astronomical Observatory, University of Warsaw, Al. Ujazdowskie 4, 00-478 Warszawa, Poland}

\author[0000-0002-2339-5899]{Pawe\l{} Pietrukowicz}
\affil{Astronomical Observatory, University of Warsaw, Al. Ujazdowskie 4, 00-478 Warszawa, Poland}

\author[0000-0003-4084-880X]{Szymon Koz\l{}owski}
\affil{Astronomical Observatory, University of Warsaw, Al. Ujazdowskie 4, 00-478 Warszawa, Poland}

\author[0000-0002-2335-1730]{Jan Skowron}
\affil{Astronomical Observatory, University of Warsaw, Al. Ujazdowskie 4, 00-478 Warszawa, Poland}

\author[0000-0002-9245-6368]{Rados\l{}aw Poleski}
\affil{Astronomical Observatory, University of Warsaw, Al. Ujazdowskie 4, 00-478 Warszawa, Poland}

\author[0000-0001-6364-408X]{Krzysztof Ulaczyk}
\affil{Department of Physics, University of Warwick, Coventry CV4 7 AL, UK}
\affil{Astronomical Observatory, University of Warsaw, Al. Ujazdowskie 4, 00-478 Warszawa, Poland}

\author[0000-0002-1650-1518]{Mariusz Gromadzki}
\affil{Astronomical Observatory, University of Warsaw, Al. Ujazdowskie 4, 00-478 Warszawa, Poland}

\author[0000-0002-9326-9329]{Krzysztof Rybicki}
\affil{Astronomical Observatory, University of Warsaw, Al. Ujazdowskie 4, 00-478 Warszawa, Poland}

\author[0000-0002-6212-7221]{Patryk Iwanek}
\affil{Astronomical Observatory, University of Warsaw, Al. Ujazdowskie 4, 00-478 Warszawa, Poland}

\author[0000-0002-3051-274X]{Marcin Wrona}
\affil{Astronomical Observatory, University of Warsaw, Al. Ujazdowskie 4, 00-478 Warszawa, Poland}


\begin{abstract}
Searches for gravitational microlensing events are traditionally concentrated on the central regions of the Galactic bulge but many microlensing events are expected to occur in the Galactic plane, far from the Galactic Center. Owing to the difficulty in conducting high-cadence observations of the Galactic plane over its vast area, which are necessary for the detection of microlensing events, their global properties were hitherto unknown. Here, we present results of the first comprehensive search for microlensing events in the Galactic plane. We searched an area of almost 3000 square degrees along the Galactic plane ($|b|<7^{\circ}$, $0^{\circ}<l<50^{\circ}$, $190^{\circ}<l<360^{\circ}$) observed by the Optical Gravitational Lensing Experiment (OGLE) during 2013--2019 and detected 630 events. We demonstrate that the mean Einstein timescales of Galactic plane microlensing events are on average three times longer than those of Galactic bulge events, with little dependence on the Galactic longitude. We also measure the microlensing optical depth and event rate as a function of Galactic longitude and demonstrate that they exponentially decrease with the angular distance from the Galactic Center (with the characteristic angular scale length of $32^{\circ}$). The average optical depth decreases from $0.5\times 10^{-6}$ at $l=10^{\circ}$ to $1.5\times 10^{-8}$ in the Galactic anticenter. We also find that the optical depth in the longitude range $240^{\circ}<l<330^{\circ}$ is asymmetric about the Galactic equator, which we interpret as a signature of the Galactic warp.
\end{abstract}


\section{Introduction} \label{sec:intro}

Gravitational microlensing events occur when an observer, a source star, and a lensing object happen to be nearly aligned so that light rays from the source are bent in the gravitational field of the lens. This results in temporary magnification of the source. Microlensing surveys are traditionally concentrated on the Galactic bulge \citep{paczynski1991,griest1991}, where the surface density of stars is the largest and so the probability of microlensing is the highest (with over 2000 events discovered annually). Although a sizable population of microlensing events is also expected in the Galactic plane fields (that is, along the Galactic equator), the lower microlensing probability, in combination with a much larger area to be surveyed, render their detection difficult. In addition, since many ``all-sky'' transient surveys avoid crowded regions of the Galactic plane by design (with a few notable exceptions), the scientifically useful sample of microlensing events in the disk is nonexistent.

As advocated by \citet{gould_lsst2013}, a microlensing survey of the Galactic plane can tackle several science questions, including studies of the Galactic distribution of exoplanets, Milky Way structure, or searches for isolated black holes, among others. From the theoretical point of view, microlensing events detected in the Galactic plane fields are expected to have longer timescales than those toward the Galactic bulge \citep{sajadian2019}, which can be attributed to the combined effects of larger Einstein radii and lower relative lens-source proper motions. Longer timescales facilitate the measurement of the annual microlens parallax effect \citep{gould1992}. Large Einstein radii increase the chances of measuring the astrometric microlensing signal \citep{hog1995,miyamoto1995,walker1995}, for example, by the \textit{Gaia} satellite. When both these effects are combined, the lens mass measurements become easier to obtain than for a typical Galactic bulge event.

\citet{gould_lsst2013} proposed carrying out a microlensing survey of the Galactic plane with the Vera C. Rubin Observatory (formerly known as the Large Synoptic Survey Telescope); also see \citet{yee2018} and \citet{street2018} for a detailed science motivation. Simulations of \citet{sajadian2019}, based on the Besan\c{c}on model of the Milky Way \citep{robin2003,robin2012}, predict that the Rubin Observatory should detect of the order of 15 microlensing events per square degree per year in the Galactic disk fields. The Milky Way model used in these simulations is well tested against the microlensing event rates and optical depths in the Galactic bulge \citep{mroz2019b} but not at large Galactic longitudes.

Only a handful of microlensing events outside the Galactic bulge were reported thus far, but these numbers are growing as new surveys dare to look at the Galactic plane. The EROS survey \citep{derue1999,derue2001,rahal2009}, as part of the seven year long campaign, found 27 microlensing event candidates toward four directions in the Galactic plane (toward Galactic longitudes of $19^{\circ}$, $27^{\circ}$, $307^{\circ}$, and $331^{\circ}$). Microlensing optical depths and timescales of the detected events are in a reasonable agreement with simple Galactic models, but owing to a small sample size, uncertainties of these quantities are large \citep{moniez2017}. \citet{fukui2007} and \citet{gaudi_halloween2008} presented the discovery of a bright microlensing event in the constellation Cassiopeia ($l=117^{\circ}$). \citet{nucita2018}, \citet{fukui2019}, and \citet{zang2019} reported the characterization of a planetary-mass companion in microlensing event TCP\,J05074264+2447555 that was located toward the Galactic anticenter ($l=179^{\circ}$). This was the first event, in which the two images generated by microlensing were resolved thanks to the use of the Very Large Telescope Interferometer GRAVITY \citep{dong2019}. This event was serendipitously discovered by T. Kojima, the discovery was reported to CBAT ``Transient Object Followup Reports.''\footnote{\url{http://www.cbat.eps.harvard.edu/unconf/followups/J05074264+2447555.html}} \citet{asassn2018} presented the discovery of two bright microlensing events in data from the ASAS-SN survey \citep{shappee2014}, one of which was located toward the Galactic anticenter ($l=190^{\circ})$. A reddened binary microlensing event PGIR~19btb ($l=54^{\circ}$) was found by the Palomar Gattini-IR survey \citep{de2019,de2020}. About 200 microlensing event candidates were discovered as part of \textit{Gaia} Science Alerts \citep[][\L{}ukasz Wyrzykowski, priv. comm.]{hodgkin2013,wyrzyk2016,kruszynska2018}, including a spectacular binary event, Gaia16aye \citep{wyrzykowski2020} ($l=65^{\circ}$). Recently, \citet{mroz2020} presented the discovery of 30 microlensing events detected in the first year of northern Galactic plane survey by the Zwicky Transient Facility \citep[ZTF;][]{bellm2019,graham2019}.

The Optical Gravitational Lensing Experiment \citep[OGLE;][]{udalski2015} is renowned for its long-term monitoring of the Galactic bulge. However, in 2013, the OGLE collaboration started the Galaxy Variability Survey (GVS), with the main goal of carrying out a variability census of the Milky Way in fields located along the Galactic plane ($|b|<7^{\circ}$, $0^{\circ}<l<50^{\circ}$, $190^{\circ}<l<360^{\circ}$) and in an extended area around the outer Galactic bulge. The fields analyzed in the current paper are presented in Figure~\ref{fig:fields}. They cover an area of about 2800~deg$^2$ and contain over 1.8~billion detected sources.

Here, we present results of the search for microlensing events in the photometric data collected as part of the OGLE GVS survey during 2013--2019. For the first time, we are able to calculate microlensing optical depth and event rate in a vast area along the Galactic plane. The methodology and structure of the paper are similar to those of our earlier work on the microlensing optical depth and event rate in the Galactic bulge \citep{mroz2019b}. The data set used in the analysis is described in Section~\ref{sec:data}. Section~\ref{sec:events} presents the selection process of microlensing events from two billions of light curves observed by the OGLE GVS. In Section~\ref{sec:stars}, we estimate the number of source stars observed in our experiment. Calculations of detection efficiency of microlensing events are summarized in Section~\ref{sec:simuls}. Finally, the main scientific results of the paper -- including measurements of the microlensing optical depth and event rate -- are presented and discussed in detail in Section~\ref{sec:results}.

\section{Data}
\label{sec:data}

The OGLE GVS survey is conducted as part of the OGLE-IV project \citep{udalski2015} using the 1.3 m Warsaw telescope located at Las Campanas Observatory, Chile. (The Observatory is operated by the Carnegie Institution for Science.) The telescope is equipped with a mosaic large-field-of-view CCD camera, which enables imaging an area of 1.4\,deg$^2$ in one exposure, with a pixel scale of $0.26''$\,pixel$^{-1}$. 

All analyzed data were taken in the $I$ band, closely matching that of a standard Cousins system. Images collected as part of the OGLE GVS are shallower than those in the Galactic bulge survey. The typical exposure times are 25 and 30\,s and the typical limiting magnitude on individual frames is $I=19$. However, we can detect microlensing events with source stars fainter than that limit. The overall depth of the survey is $I=21$ (Sections \ref{sec:stars} and \ref{sec:results}). The OGLE photometric pipeline is based on the \citeauthor{wozniak2000}'s (\citeyear{wozniak2000}) implementation of the difference image analysis (DIA) technique \citep{alard1998}. For each field, a reference image is constructed by stacking 3--14 good-seeing low-background individual images. Subsequently, the reference image is subtracted from the incoming images and the photometry is performed on the subtracted images. The DIA technique was specifically developed for extremely crowded sky areas, such as the central regions of Galactic bulge. The stellar density in the presently analyzed fields is 1--2 orders of magnitude smaller than that in the Galactic bulge. Nonetheless, the use of DIA enables us to achieve the photometric precision of 2--3~mmags for bright sources. The photometry is calibrated to the standard system, and the accuracy of the zero-point transformations is 0.01--0.02 mag. 

The photometric data are stored in two separate databases. The ``standard'' database contains measurements for all stellar objects detected in the reference image, and the photometry of ``new'' objects (that is, not detected on the reference image) is stored separately. In this paper, we analyze photometry from the ``standard'' database. 

The photometric uncertainties returned by DIA are known to be usually underestimated \citep[e.g.,][]{yee2012}. Thus, we used the method developed by \citet{skowron2016} to correct the reported uncertainties, so that they reflect the actual scatter in the data. Each error bar was rescaled using the formula $\delta m_{i,\mathrm{new}}=\sqrt{(\gamma\delta m_i)^2+\varepsilon^2}$, where the coefficients $\gamma$ and $\varepsilon$ were determined for each field and CCD detector separately. Their typical values are $\gamma=1.2-1.4$ and $\varepsilon=0.003-0.005$. For the brightest stars ($I\lesssim 14$), an additional correction due to the nonlinearity of the detector must be applied \citep{skowron2016}.

\begin{sidewaysfigure}
\includegraphics[width=\textheight]{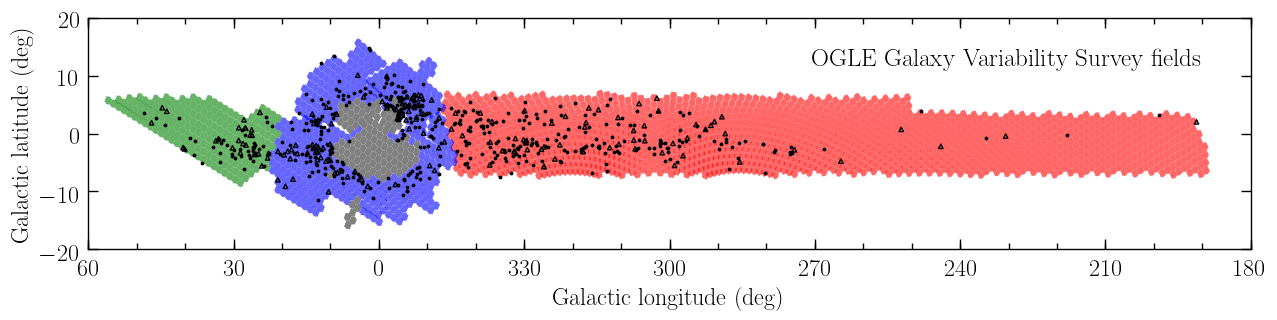}\\
\includegraphics[width=\textheight]{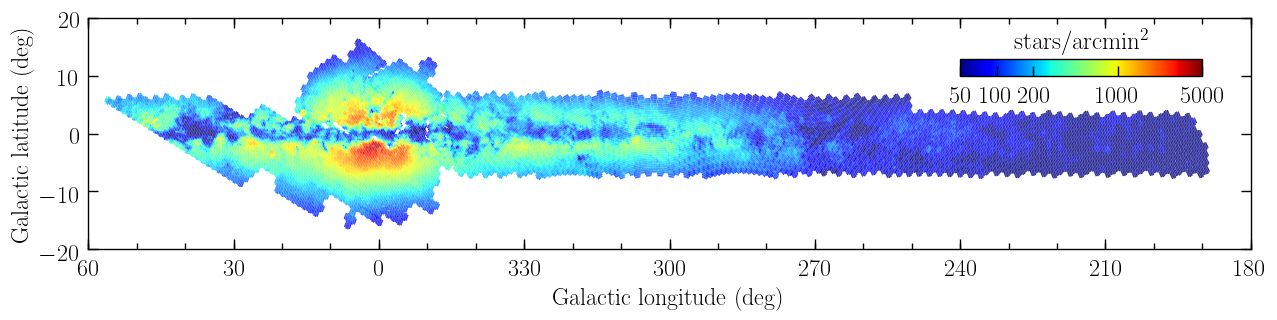}
\caption{Upper panel: fields of the OGLE Galaxy Variability Survey are marked in red (GD), green (DG), and blue (BLG). Previously analyzed fields covering the central regions of the Galactic bulge \citep{mroz2019b} are marked in dark gray. Filled circles mark microlensing events used for the optical depth and event rate measurements, whereas empty triangles mark possible microlensing events (see Section~\ref{sec:events}). Lower panel: surface density of source stars brighter than $I=21$ in OGLE fields.}
\label{fig:fields}
\end{sidewaysfigure}

The footprint of the survey is presented in the upper panel of Figure~\ref{fig:fields}. The currently analyzed area consists of 1338 Galactic disk fields located west of the Galactic Center ($190^{\circ}<l<345^{\circ}$), 200 Galactic disk fields located east of the Galactic Center ($20^{\circ}<l<50^{\circ}$), and 444 fields covering the outer Galactic bulge. These fields are named GDNNNN, DGNNNN, and BLGNNN, respectively, where NNNN (NNN) is a four digit (three digit) number. Basic information about the analyzed fields (their coordinates, number of sources in the database, and number of epochs) is presented in Table~\ref{tab:allfields}.

\begin{figure}
\includegraphics[width=\textwidth]{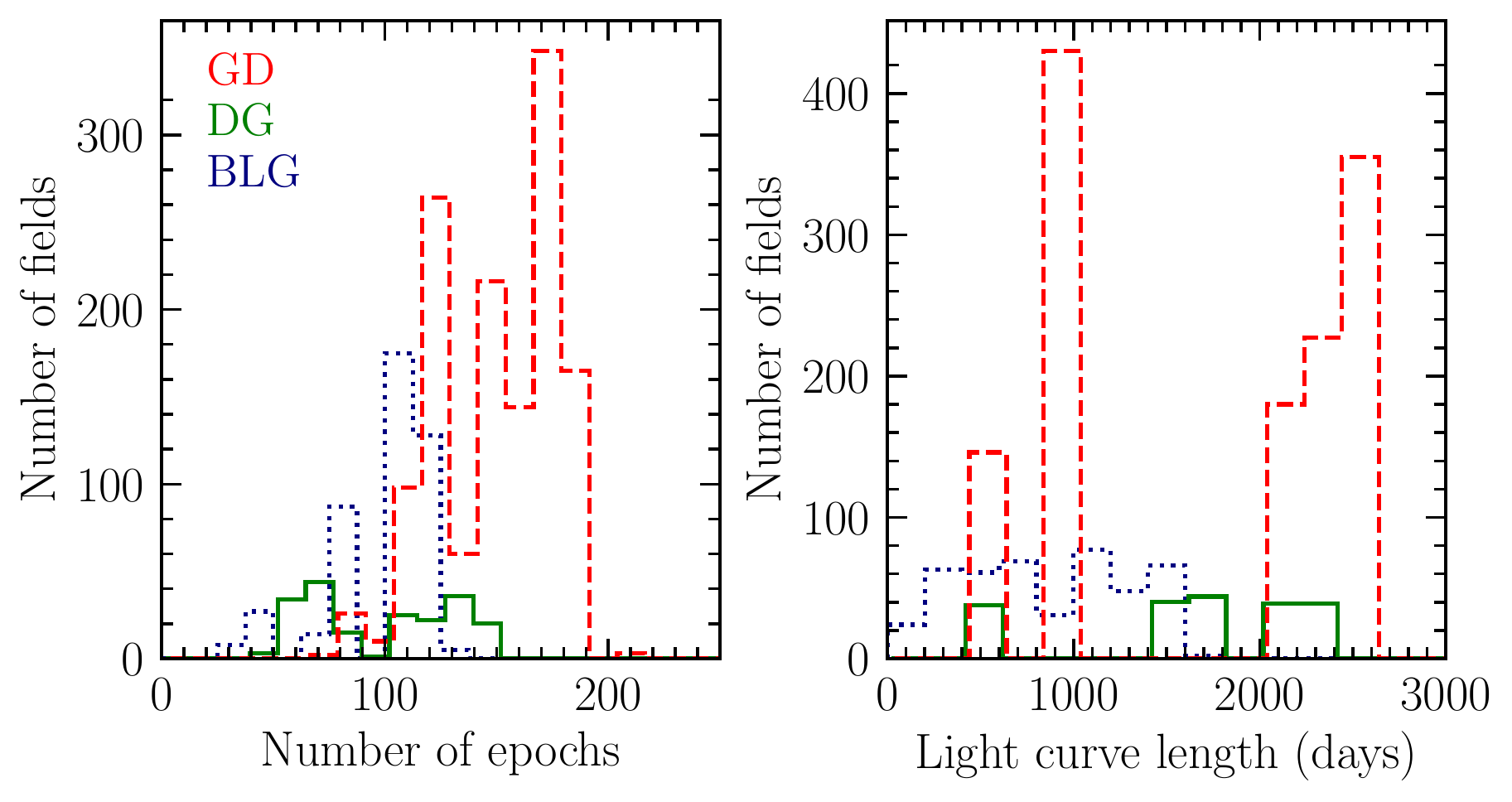}
\caption{Number of epochs and duration of monitoring of the OGLE GVS fields (GD ($190^{\circ}<l<345^{\circ}$): dashed red, DG ($20^{\circ}<l<50^{\circ}$): solid green line, BLG (outer bulge): dotted blue line).}
\label{fig:stats}
\end{figure}

The number of epochs and the duration of monitoring vary depending on the field. As presented in Figure~\ref{fig:stats}, most of the analyzed fields were observed 100--200 times during a period of 2--7~yr. Only 13\% of the fields have fewer than 100 epochs and 19\% have been monitored for less than 2~yr. The sampling of the light curves is not uniform. Usually, fields were observed with a 1--2 day cadence during one or two observing seasons, whereas the remaining part of the light curve is more sparsely sampled (with observations 10--15 days apart).

\section{Selection of microlensing events}
\label{sec:events}

Gravitational microlensing leads to a temporary, non-repeatable brightening of the source star. Microlensing events caused by single lenses have characteristic symmetrical light curves with the wide range of durations (days to weeks to years) and amplitudes (from submagnitude level to 5~mag and above). There are, however, many other astrophysical transient sources in the Galactic plane, such as flaring stars, dwarf novae and other cataclysmic variable stars, X-ray binary systems, young stellar objects, Be-type stars, etc. Our selection method is devised to remove astrophysical (and instrumental) contamination, while retaining genuine microlensing events.

The search method and final selection criteria are similar to those used in our previous works \citep[][which are themselves based on \citealt{sumi2011}]{mroz2017,mroz2019b}, although with some small changes. These changes are introduced to compensate for a sparser coverage of the light curves as compared to high-cadence observations of the Galactic bulge. Our selection process consists of three steps: selecting transient objects, removing obvious non-microlensing light curves, and fitting microlensing model to the data. All selection criteria are applied in an automated fashion. They are summarized in Table~\ref{tab:crit}.

In the first step, we search for light curves that exhibit a transient brightening and do not show any additional variability. We place a moving window on the light curve and calculate the mean flux ($F_{\rm base}$) and its dispersion ($\sigma_{\rm base}$) outside the window (we use $5\sigma$ clipping to remove obvious outliers). To quantify the variability of the source, we calculate $\chi^2_{\rm out}=\sum_i (F_i-F_{\rm base})^2/\sigma^2_i$ (the summation is performed over all $N$ data points outside the window) and we require $\chi^2_{\rm out}\leq 2N$. Here ($F_i$ and $\sigma_i$) represent the flux and its uncertainty of the $i$th data point.
As a rule of thumb, the length of the window should be shorter than $\sim 30-40\%$ of the light-curve length. This makes sure that there is a sufficient number of data points outside the window to detect any variability in the baseline. Here, we use a 720\,day window for fields observed for longer than 2000\,days, 360\,day for fields observed for longer than 900\,days, and 120\,day for the remaining fields.

Subsequently, we search for at least three consecutive data points within the window  fulfilling the condition $F_i \geq F_{\rm base}+3\sigma_{\rm base}$ and calculate $\chi_{3+}=\sum_i(F_i-F_{\rm base})/\sigma_i$ for them. We require $\chi_{3+}\geq 32$. We also require that the additional light on the subtracted image was indeed clearly above the noise and assigned to the object at hand (and not to one of its close neighbors) at least three times during these consecutive brightened points ($n_{\rm DIA} \geq 3$).

In the next step, we reject light curves with more than one brightening (such as dwarf novae) and with low-amplitude brightenings (these are mostly pulsating red giants). These typically would be rejected by the latter requirements on the goodness of the microlensing fit, but it is prudent to remove them earlier.

Finally, we fit a point-lens point-source (PSPL) microlensing model to the remaining light curves by minimizing $\chi^2_{\rm fit} = \sum_i (F_i-F_{i,\mathrm{model}})^2/\sigma^2_i$ using a downhill algorithm, where $F_{i,\mathrm{model}}=F_{\rm s} A(t_i) + F_{\rm b}$, $\Fs$ and $\Fb$ are source and blend fluxes, respectively, and:
\begin{align*}
A(t) &= \frac{u^2(t)+2}{u(t)\sqrt{u^2(t)+4}}, \\
u(t) &= \sqrt{u_0^2+\left(\frac{t-t_0}{\tE}\right)^2}.
\end{align*}
Here ($t_0,u_0,\tE$) are principal microlensing model parameters, the time of and separation during the closest approach between the lens and the source, and the Einstein radius crossing timescale. We require $\chi^2/\mathrm{dof}\leq 2$ for the entire light curve and for data points centered on the peak ($|t_i-t_0|<\tE$). We use additional cuts on the values of $t_0$, $u_0$, $\tE$, $\Fs$, and $\Fb$ (Table~\ref{tab:crit}). We require long-timescale events ($\tE \geq$\,100\,days) to have amplitudes higher than 0.4\,mag to remove contamination from outbursts of Be-type stars, which often exhibit low-amplitude long-timescale variability \citep{mennickent2002}. 

We also require the Einstein timescale to be ``well-measured.'' We fit PSPL models with Einstein timescales fixed to $0.5\tE$ and $2\tE$ (where $\tE$ is the best-fitting timescale) and require that their $\chi^2$s are larger than $\chi^2_{\rm fit}+1$. This ensures that the fractional uncertainty of $\tE$ is lower than $50\%$ (in practice, timescales are measured with a better precision).

Our final sample of microlensing events in the OGLE GVS fields comprises 460 objects, which are listed in Table~\ref{tab:params}, together with the best-fit model parameters and their uncertainties. The uncertainties were calculated using the Markov Chain Monte Carlo sampler by \citet{foreman2013} in which we assumed the following prior on the blend flux:
\begin{equation}
\mathcal{L}_{\rm prior} = 
\begin{cases}
1 & \mathrm{if}\ F_{\rm b} \geq 0, \\
\exp{\left(-\frac{F_{\rm b}^2}{2\sigma^2}\right)} & \mathrm{if}\ F_{\rm b} < 0,
\end{cases}
\end{equation}
where $\sigma=F_{\rm min}/3$ and $F_{\rm min}$ is the flux corresponding to $I=20.5$. The uncertainties represent the 68\% confidence range of the marginalized posterior distribution. Einstein timescales of selected events are relatively well measured with the median fractional uncertainty of $\sigma(\tE)/\tE = 18\%$.

We visually inspected the light curves of transient objects that did not meet our selection criteria and identified 170 additional possible microlensing events. Many of these objects have poorly sampled or noisy light curves, so they cannot be securely classified as microlensing events. Of the 170 possible microlensing events, 63 (or 37\%) did not pass our goodness of the fit criteria ($\chi^2_{\rm fit}/\rm{dof} \leq 2$ and $\chi^2_{\rm fit,\tE}/\rm{dof} \leq 2$), including 30 events with signatures of binary lens and 3 events with a strong annual parallax effect (GD1298.15.107/Gaia19bld, GD1326.12.22665, and GD2020.09.329/Gaia19aqw). Of the events, 68 (40\%) were rejected because their Einstein timescales were poorly measured (as explained above), whereas 39 events (23\%) were missed for various other reasons (for example, they peaked before (after) the start (end) of observations). For the sake of completeness, the coordinates of possible microlensing events are listed in Table~\ref{tab:possible}. These events are not, however, used in the statistical analysis.

We crosscorrelated our lists of microlensing events with objects reported in the Transient Name Server\footnote{\url{https://wis-tns.weizmann.ac.il/}}, finding that 23 of them were previously reported as possible astrophysical transients by the \textit{Gaia} Science Alerts program\footnote{\url{http://gsaweb.ast.cam.ac.uk/alerts}} \citep{hodgkin2013}. The All-Sky Automated Survey for SuperNovae \citep{shappee2014} detected four events and two were found by \citet{mroz2020} in the photometric data from ZTF.

\begin{table}[b]
\caption{Selection Criteria for High-quality Microlensing Events in OGLE GVS Fields.}
\label{tab:crit}
\centering
\footnotesize
\begin{tabular}{p{0.3\textwidth}p{0.48\textwidth}p{0.12\textwidth}}
\hline
Criteria & Remarks & Number \\
\hline
All stars in databases & & 1,856,529,265\\
\hline
$\chi^2_{\rm out}/{\rm dof} \leq 2.0$ & No variability outside a window centered \mbox
{on the event} (duration of the window depends on the field)\\
$n_{\rm DIA} \geq 3$ & Centroid of the additional flux coincides with the source star centroid\\
$\chi_{3+}=\sum_i(F_i-F_{\rm base})/\sigma_i\geq 32$ & Significance of the bump & 23,618\\
\hline
$A \geq 0.1$ mag & Rejecting low-amplitude variables \\
$n_{\rm bump}=1$ & Rejecting objects with multiple bumps & 18,397 \\
\hline
& Fit quality: & \\
$\chi^2_{\rm fit}/\rm{dof} \leq 2.0$ & $\chi^2$ for all data \\
$\chi^2_{\rm fit,\tE}/\rm{dof} \leq 2.0$ & $\chi^2$ for $|t-t_0|<\tE$ \\
$\sigma(\tE)/\tE < 0.5$ & Einstein timescale is well measured\\
$t_{\rm min}\leq t_0\leq t_{\rm max}$ & Event peaked between $t_{\rm min}$ and $t_{\rm max}$, which are moments of the first and last observation of a given field \\
$u_0 \leq 1$ & Maximum impact parameter \\
$\tE \leq 500$\,d & Maximum timescale \\
$A\geq 0.4$\,mag if $\tE \geq 100$\,days & Long-timescale events should have high amplitudes \\
$I_{\rm s} \leq 21.0$ & Maximum $I$-band source magnitude \\
$F_{\rm b} > -F_{\rm min}$ &  Maximum negative blend flux, corresponding \mbox{to $I=20.5$ mag star}  & 460\\
\hline
\end{tabular}
\end{table}

\section{Star counts}
\label{sec:stars}

The number of monitored sources is an essential quantity needed for the calculation of the microlensing optical depth and event rate. The typical star surface density in the OGLE GVS fields is 1--2 orders of magnitude smaller than that in the densest regions of the Galactic bulge, which -- in combination with the pixel size of the OGLE-IV camera ($0.26''$\,pixel$^{-1}$) and typical PSF size on the reference images ($<1''$) -- make the OGLE star catalogs highly complete. Nonetheless, we run extensive image-level simulations to measure the completeness of stars counts as a function of magnitude and to derive the distribution of the blending parameter in all analyzed fields.

Simulations were carried out in a similar fashion to our previous study \citep{mroz2019b}. We injected 5000 artificial stars (625 were drawn uniformly from the magnitude range of $14<I<18$, 1875 from the range $18<I<20$, and 2500 from the range $20<I<21$) into images of a given subfield using random locations and appropriate point-spread function. Then, we combined them to create a deep reference image and constructed star catalogs using the same pipeline as that used to create real photometric maps \citep{udalski2003}. The main goal of the simulations was to check if the injected star was detected by our star detection pipeline (as explained in Section~4 of \citealt{mroz2019b} in detail), and, if yes, to check if it is blended with another star that was present in the original image. 

We created 126,848 artificial reference images (two for each of 1982 fields, where each field consists of 32 subfields), where each reference image is composed of 3--14 individual frames. Simulations were run on a cluster of about 400 modern CPUs for about a week ($\sim 8$ CPU years), each simulation (injecting stars into individual frames and creating reference image and the star list) took up about 30~minutes.

\begin{figure}
\centering
\includegraphics[width=\textwidth]{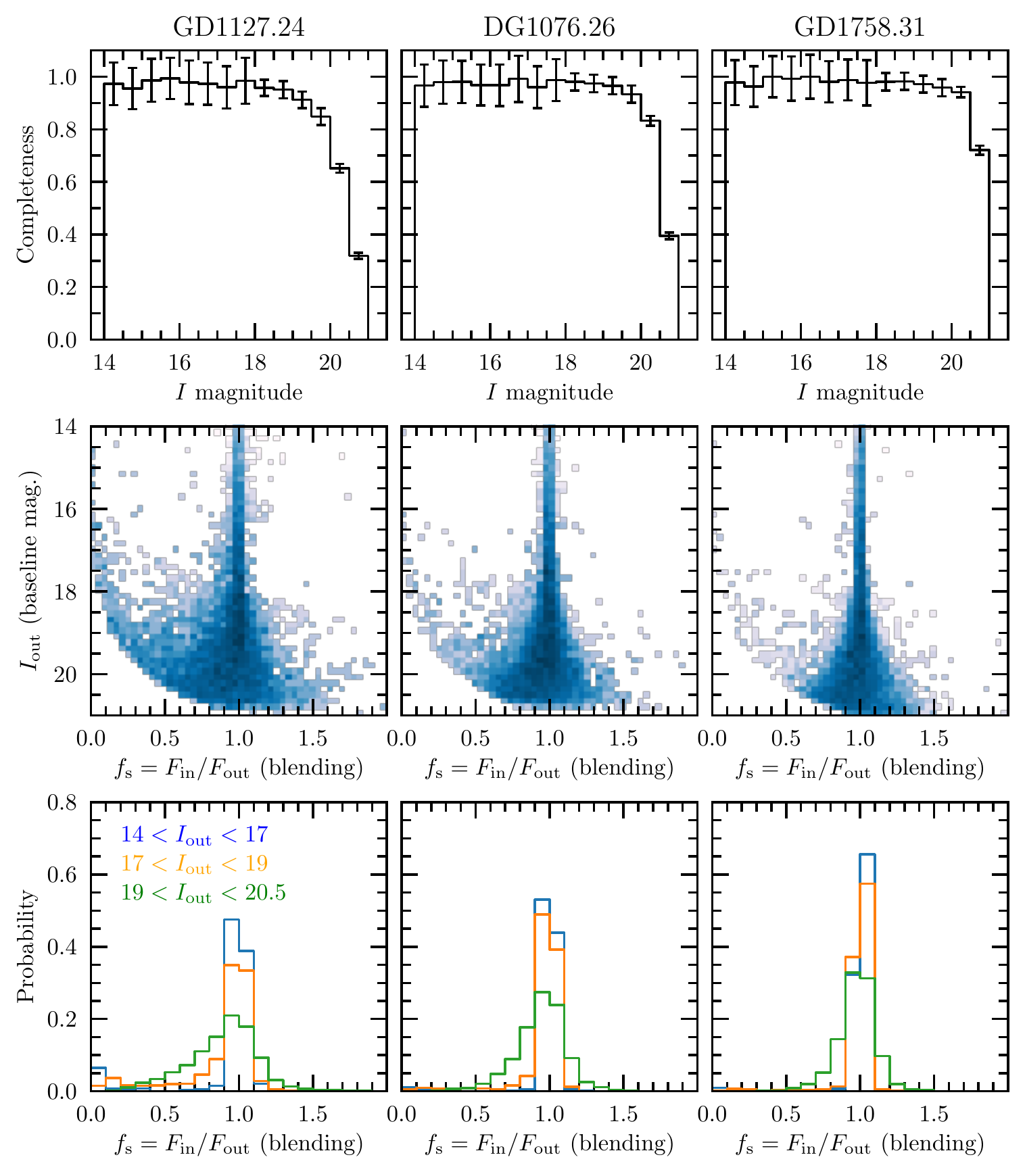}
\caption{Results of image-level simulations in three representative fields of the OGLE GVS: GD1127.24 (603.6 stars\,arcmin$^{-2}$), DG1076.26 (133.7 stars\,arcmin$^{-2}$), and GD1758.31 (58.1 stars\,arcmin$^{-2}$). Typical star surface densities are 1--2 orders of magnitude smaller than in the Galactic bulge fields \citep{mroz2019b}. Top panels: completeness of OGLE star counts as a function of magnitude. Middle and bottom: distributions of the dimensionless blending parameter.}
\label{fig:simuls}
\end{figure}

Results of the simulations are summarized in Figure~\ref{fig:simuls}. Upper panels present the completeness of star counts as a function of magnitude in three representative fields. Typical exposure times in the OGLE GVS survey ($25-30$\,s) are shorter than in the regular survey ($100-110$\,s), but the reference images are composed of a larger number of individual frames, making it possible to detect stars as faint as $I\approx 21$. Star catalogs are nearly complete down to $I=20$ (the median completeness is 97\%) and the completeness remains high down to $I=21$ (with the median value of 85\%).

The middle and bottom panels of Figure~\ref{fig:simuls} present the distribution of the dimensionless blending parameter as a function of the baseline magnitude. They were created by matching stars injected into images with those found on the simulated reference image. The blending parameter is simply $f_{\rm s}=F_{\rm in}/F_{\rm out}$, where $F_{\rm out}$ is the flux measured on the reference image and $F_{\rm in}$ is the simulated flux. For the majority of simulated stars, $f_{\rm s}\approx 1$, indicating little or no blending. The inferred blending distributions significantly differ from those in the Galactic bulge (Figure~7 of \citealt{mroz2019b}), where the surface density of stars is 1--2 orders of magnitude higher. For example, there are very few source stars with $f_{\rm s}\approx 0$. The results of our image-level simulations also indicate that whenever $f_{\rm s} \neq 1$ is measured from the light curve, the blended light is very likely to originate from the lens itself.

\begin{figure}
\centering
\includegraphics[width=0.7\textwidth]{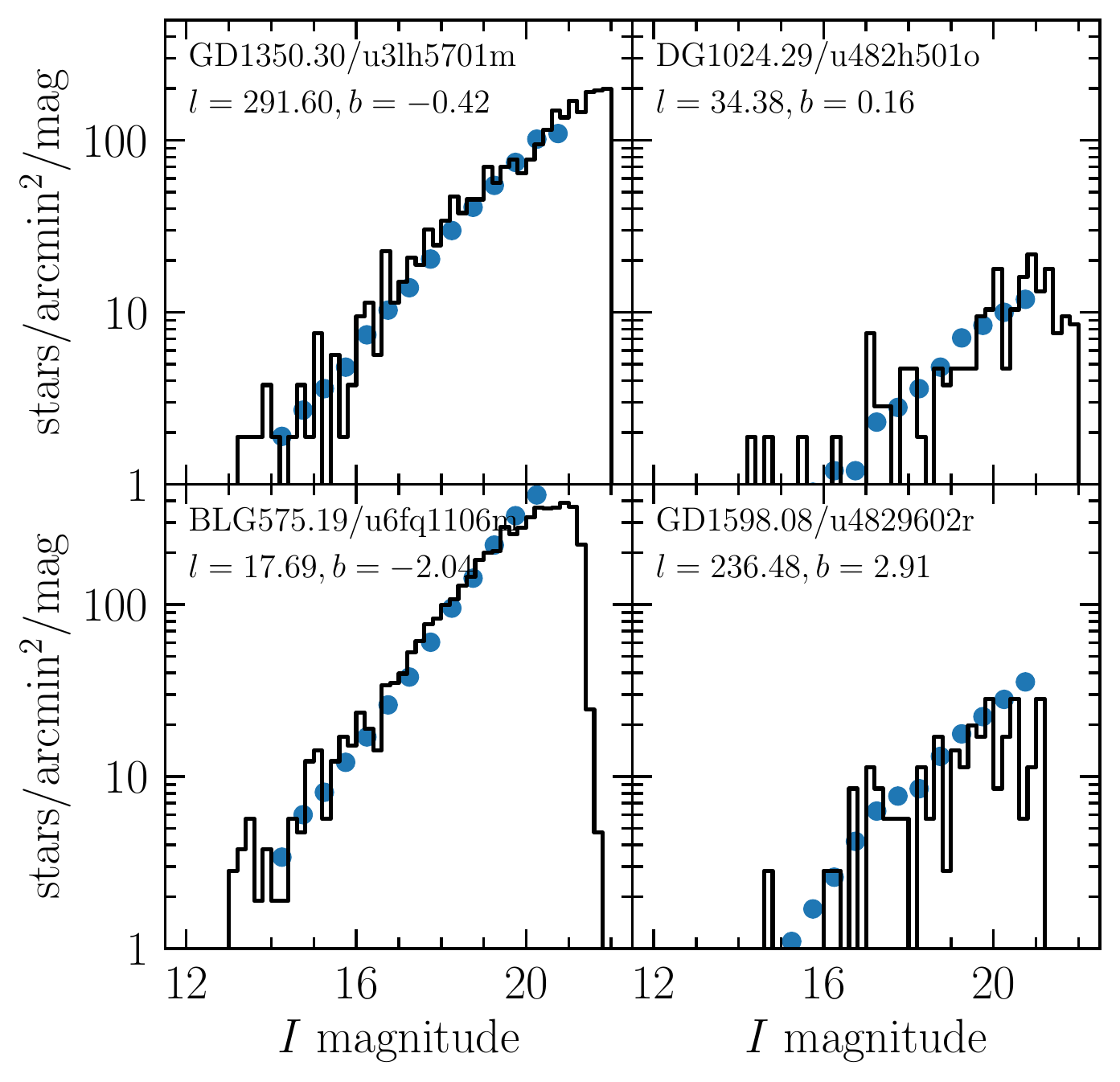}
\caption{Comparison between our completeness-corrected stellar luminosity functions (blue dots) and those derived from \textit{HST} images (black histogram). Names and Galactic coordinates of analyzed fields are in the upper left corner. \textit{HST} images u6fq1106m and u4829602r (lower panels) are incomplete for $I\gtrsim 20$ because of short exposure times (60 and 40~s, respectively).}
\label{fig:hst}
\end{figure}

We used our image-level simulations to estimate the completeness of star counts as a function of magnitude. That allowed us to correct the observed luminosity functions for incompleteness in each of the analyzed fields. (As discussed above, these corrections were usually small.) The measured completeness-corrected surface densities of stars (down to $I=20$ and $I=21$) in all analyzed fields are presented in Table~\ref{tab:stars}. Additionally, the lower panel of Figure~\ref{fig:fields} presents the surface density of stars brighter than $I=21$ in the OGLE GVS fields and in the previously analyzed Galactic bulge fields \citep{mroz2019b}.

To validate results of our calculations, we analyzed images of the four Galactic plane fields collected using the Wide Field Planetary Camera 2 (WFPC2) on board the \textit{Hubble Space Telescope} \citep[\textit{HST};][]{holtzman2006}, which has a pixel scale of $\sim 0.1''$\,pixel$^{-1}$. We aimed to derive stellar luminosity functions (in absolute units) in each \textit{HST} image and compare them with the completeness-corrected luminosity functions based on OGLE data. The selected images were taken through the F814W filter and calibrated against OGLE photometric maps. The \textit{HST} photometry was obtained with the Dolphot package\footnote{http://americano.dolphinsim.com/dolphot/} \citep{dolphin2000}. The image-to-sky coordinate transformations were done using the WCSTools package \citep{mink1997}. Figure~\ref{fig:hst} presents the comparison between completeness-corrected luminosity functions based on OGLE images and those measured using the \textit{HST} data. Both luminosity functions agree well in all four \textit{HST} fields.

\section{Detection efficiency simulations}
\label{sec:simuls}

We carried out extensive catalog-level simulations to measure the detection efficiency of microlensing events as a function of their timescales. We generated synthetic light curves of microlensing events by injecting a microlensing model on top of the light curves of objects from the OGLE GVS databases. Technical details of simulations are described in Section~6 of \citet{mroz2019b}. In short, each data point and its error bar were rescaled in accordance to the expected magnification, which depends on the microlensing model and blending parameter. Thus, our method conserves noise and variability in the original light curves, as well as information about the quality of individual measurements, which would be otherwise difficult to simulate.
In our previous work \citep{mroz2019b}, we demonstrated that catalog-level simulations provide nearly identical results to those of more resource-consuming image-level simulations (in which artificial microlensing events are injected into CCD images).

\begin{figure}
\centering
\includegraphics[width=0.7\textwidth]{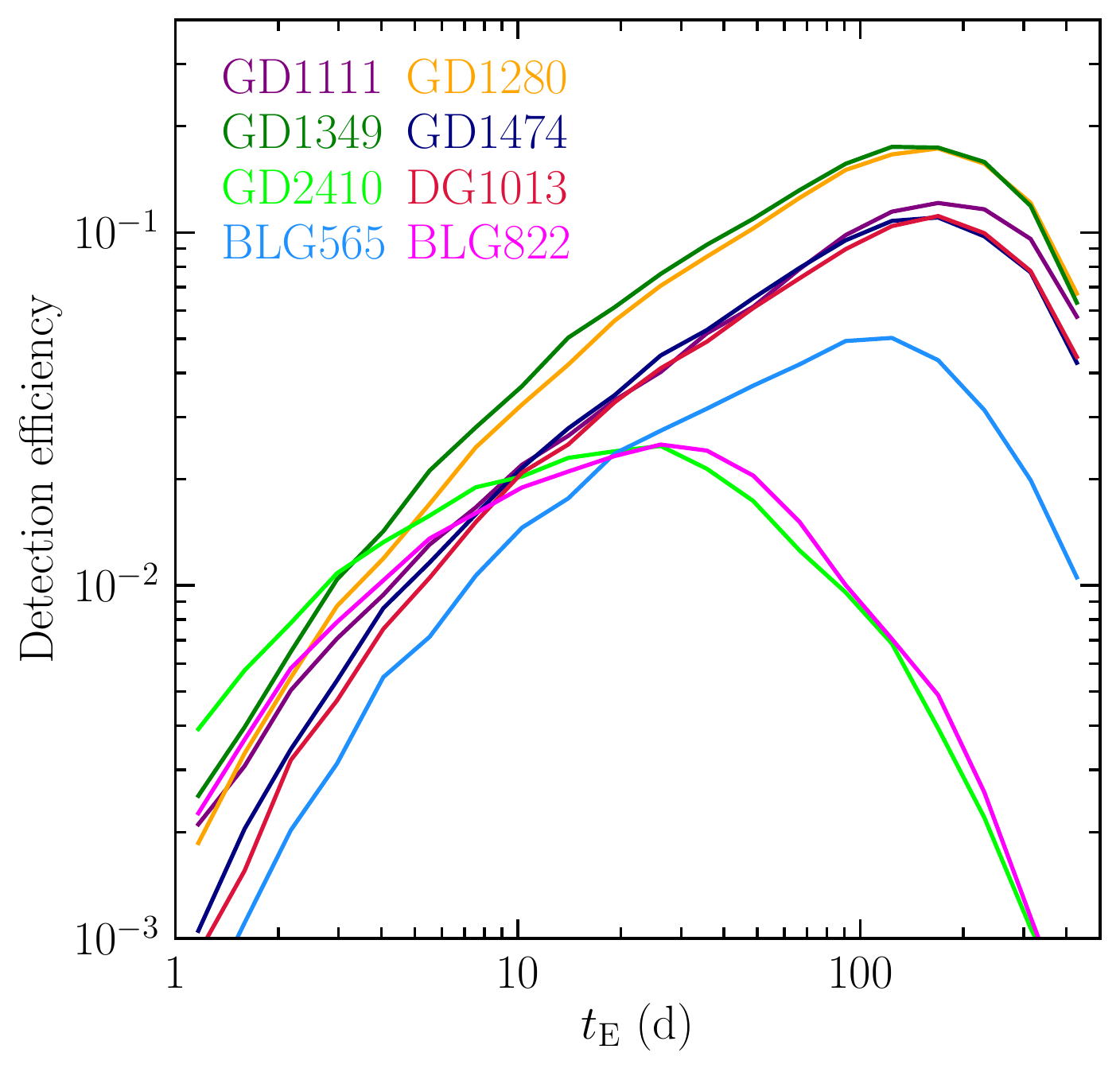}
\caption{Examples of detection efficiency curves. Fields GD2410 and BLG822 have been observed for less than 500\,days and so sensitivity to long-timescale events in these fields is low. The detection efficiency curves have different normalizations, because the analyzed fields were observed for different lengths of time. (The detection efficiencies measured in our simulations have been multiplied by a factor of $(t_{\rm max}-t_{\rm min})/\Delta T$, where $t_{\rm min}$ and $t_{\rm max}$ are moments of the first and last observation of a given field and $\Delta T =2650\,\mathrm{days}$ is the duration of the survey.)}
\label{fig:eff}
\end{figure}

We simulated 50,000 events for each CCD detector -- that is, 1.6~million light curves per field ($\sim 3$ billion events in total). Simulations were run on a cluster of about 50 modern CPUs for about a day (in total, $\sim 60$ CPU days). The parameters of the simulated events were randomly drawn from uniform distributions: $t_0\sim U(t_{\rm min},t_{\rm max})$, $u_0\sim U(0,1)$, and $\log\tE \sim U(0.0,2.7)$, where $t_{\rm min}$ and $t_{\rm max}$ are moments of the first and last observation of a given field. Subsequently, we drew a random star from the database and calculated its mean magnitude. The blending parameter $f_{\rm s}$ was randomly selected from the empirical distribution derived using our image-level simulations (Section~\ref{sec:stars}) based on the brightness of the baseline object. Then, we generated synthetic light curves and checked if they pass our selection criteria (Table~\ref{tab:crit}). The analyzed fields were observed for different lengths of time, so the detection efficiencies measured in our simulations have been multiplied by a factor of $(t_{\rm max}-t_{\rm min})/\Delta T$, where $\Delta T =2650\,\mathrm{days}$ is the duration of the survey (between 2012 September 29 and 2020 January 1). Thus, detection efficiency curves presented in Figure~\ref{fig:eff} and in the online data represent the probability of finding a microlensing event with a source brighter than $I=21$ during a period of 2650\,days.

\section{Results}
\label{sec:results}

\subsection{Microlensing optical depth and event rate}

The optical depth to gravitational microlensing is defined as the fraction of sky covered by Einstein rings of lensing objects. It is thus proportional to the fraction of time sources spend inside the Einstein ring:
\begin{equation}
\tau = \frac{\pi}{2N_{\rm s}\Delta T}\sum_i\frac{t_{{\rm E},i}}{\varepsilon(t_{{\rm E},i})},
\end{equation}
where $t_{{\rm E},i}$ is the Einstein timescale of the $i$th event, $\varepsilon(t_{{\rm E},i})$ is the event detection efficiency at that timescale, $N_{\rm s}$ is the number of monitored source stars, and $\Delta T$ is the duration of the experiment. Here, we use $\Delta T=2650\,\mathrm{days}\approx 7.3\,\mathrm{yr}$. (We note that the analyzed fields were observed for different lengths of time, but the detection efficiencies $\varepsilon(t_{{\rm E},i})$ have been rescaled to match the duration of the survey; see Section~\ref{sec:simuls}.)

Similarly, the event rate per source star is given by
\begin{equation}
\Gamma = \frac{1}{N_{\rm s}\Delta T}\sum_i\frac{1}{\varepsilon(t_{{\rm E},i})}.
\end{equation}
We also calculate the event rate per unit area, which does not explicitly depend on the star counts:
\begin{equation}
\Gamma_{\rm deg^2} = \frac{1}{\Delta\Omega\Delta T}\sum_i\frac{1}{\varepsilon(t_{{\rm E},i})},
\end{equation}
where $\Delta\Omega$ is the analyzed area. The uncertainties are measured using formulae of \citet{han1995_stat} and \citet{mroz2019b}. We note that the measured optical depths and event rates are averaged over all source stars brighter than $I=21$.

A small fraction of all events ($\approx 10\%$) are anomalous (for example, binary lens events). These events were detected by our search algorithm but were rejected by the cuts on the goodness of the microlensing PSPL fit, so the measured microlensing optical depth and event rate may be slightly underestimated. Thus, following \citet{sumi2013} and \citet{mroz2019b}, we rescale optical depths and event rates (and their uncertainties) by a factor of 1.09. (Binary microlensing events identified during the visual inspection of light curves are listed in Table~\ref{tab:possible}.) Similarly, long-timescale events exhibiting the strong annual parallax effect were excluded from the final sample (we found three such events; they are listed in Table~\ref{tab:possible}). Other possible sources of systematic errors (for example, neglecting ``new object'' channel events in the catalog-level simulations and using catalog-level simulations instead of image-level simulations for measuring detection efficiencies) are negligible, as discussed in detail by \citet{mroz2019b}. We also note that we use events with timescales shorter than $\tE=500$\,days for the measurements of the optical depth and event rates but, in principle, longer-timescale events may occur. Thus, we add a subscript ``500'' to the measured values of $\tau$ to emphasize that such a limit is in place.

\begin{deluxetable}{lrrrrrr}[h]
\tablecaption{Microlensing Optical Depth and Event Rate toward the Galactic Plane (Averaged over Sources Brighter than $I_{\rm s}=21$) \label{tab:tau}}
\tablehead{
\colhead{Region} & \colhead{$N_{\rm ev}$} & \colhead{$N_{\rm s}$} & \colhead{$\tau_{500}$} &
\colhead{$\Gamma$} & \colhead{$\Gamma_{\rm deg^2}$} & \colhead{$\langle\tE\rangle$} \\
\colhead{} & \colhead{} & \colhead{($10^6$)} & \colhead{($10^{-6}$)} &
\colhead{($10^{-6}\,\mathrm{yr}^{-1}$)} & \colhead{($\mathrm{yr}^{-1}\,\mathrm{deg}^{-2}$)} & \colhead{(days)} 
}
\startdata
Galactic longitude bins: \\
$190<l<270$,$-7<b<7$ & 5 & 208.217 & $0.015 \pm 0.007$ & $0.073 \pm 0.033$ & $0.017 \pm 0.008$ & $47.4 \pm 7.9$ \\ 
$270<l<290$,$-7<b<7$ & 18 & 140.019 & $0.094 \pm 0.028$ & $0.379 \pm 0.119$ & $0.190 \pm 0.060$ & $57.9 \pm 18.2$ \\ 
$290<l<300$,$-7<b<7$ & 14 & 109.993 & $0.099 \pm 0.029$ & $0.353 \pm 0.103$ & $0.269 \pm 0.079$ & $65.1 \pm 13.7$ \\ 
$300<l<310$,$-7<b<7$ & 22 & 133.120 & $0.175 \pm 0.048$ & $0.504 \pm 0.118$ & $0.482 \pm 0.113$ & $80.5 \pm 15.6$ \\ 
$310<l<320$,$-7<b<7$ & 32 & 129.665 & $0.252 \pm 0.070$ & $0.919 \pm 0.189$ & $0.810 \pm 0.167$ & $63.8 \pm 15.3$ \\ 
$320<l<330$,$-7<b<7$ & 34 & 149.050 & $0.220 \pm 0.046$ & $0.806 \pm 0.156$ & $0.842 \pm 0.163$ & $63.5 \pm 10.1$ \\ 
$330<l<340$,$-7<b<7$ & 47 & 152.811 & $0.334 \pm 0.071$ & $1.362 \pm 0.221$ & $1.461 \pm 0.237$ & $57.0 \pm 9.9$ \\ 
$340<l<350$,$-7<b<7$ & 40 & 151.878 & $0.474 \pm 0.098$ & $2.122 \pm 0.412$ & $2.338 \pm 0.454$ & $52.0 \pm 7.7$ \\ 
$10<l<20$,$-7<b<7$ & 54 & 205.197 & $0.594 \pm 0.094$ & $3.112 \pm 0.498$ & $5.117 \pm 0.818$ & $44.4 \pm 4.6$ \\ 
$20<l<30$,$-7<b<7$ & 25 & 144.309 & $0.252 \pm 0.083$ & $0.978 \pm 0.228$ & $1.360 \pm 0.318$ & $59.9 \pm 17.8$ \\ 
$30<l<60$,$-7<b<7$ & 18 & 163.916 & $0.236 \pm 0.078$ & $0.889 \pm 0.263$ & $0.738 \pm 0.218$ & $61.7 \pm 12.8$ \\ 
\hline
Galactic latitude bins: \\
$240<l<330$,$-7<b<-5$ & 6 & 95.625 & $0.076 \pm 0.044$ & $0.387 \pm 0.176$ & $0.193 \pm 0.088$ & $45.5 \pm 25.8$ \\ 
$240<l<330$,$-5<b<-3$ & 20 & 136.125 & $0.131 \pm 0.040$ & $0.565 \pm 0.140$ & $0.398 \pm 0.098$ & $53.8 \pm 12.5$ \\ 
$240<l<330$,$-3<b<-1$ & 51 & 167.427 & $0.277 \pm 0.059$ & $0.822 \pm 0.129$ & $0.717 \pm 0.112$ & $78.3 \pm 13.7$ \\ 
$240<l<330$,$-1<b<1$ & 14 & 103.781 & $0.119 \pm 0.036$ & $0.435 \pm 0.161$ & $0.242 \pm 0.089$ & $63.5 \pm 21.2$ \\ 
$240<l<330$,$1<b<3$ & 13 & 118.778 & $0.094 \pm 0.027$ & $0.304 \pm 0.089$ & $0.197 \pm 0.058$ & $71.9 \pm 12.3$ \\ 
$240<l<330$,$3<b<5$ & 15 & 96.696 & $0.155 \pm 0.046$ & $0.597 \pm 0.163$ & $0.356 \pm 0.097$ & $60.5 \pm 12.5$ \\ 
$240<l<330$,$5<b<7$ & 3 & 50.281 & $0.080 \pm 0.048$ & $0.283 \pm 0.167$ & $0.121 \pm 0.071$ & $66.0 \pm 16.8$ \\ 
\enddata
\end{deluxetable}

The microlensing optical depth primarily depends on the mass density of lenses along the line of sight. We thus expect that the optical depth in the Galactic plane should decrease with the angular distance from the Galactic Center, as the line of sight encloses a smaller volume of the Galaxy. To measure these expected variations, we calculate the optical depth and event rate in 11 bins in Galactic longitude. Most bins extend $10^{\circ}$ along the Galactic equator and cover the latitude range $-7^{\circ}<b<7^{\circ}$ (the three outermost bins are wider, from $20^{\circ}$ to $80^{\circ}$, and have better statistics). 

\begin{figure}
\centering
\includegraphics[width=.49\textwidth]{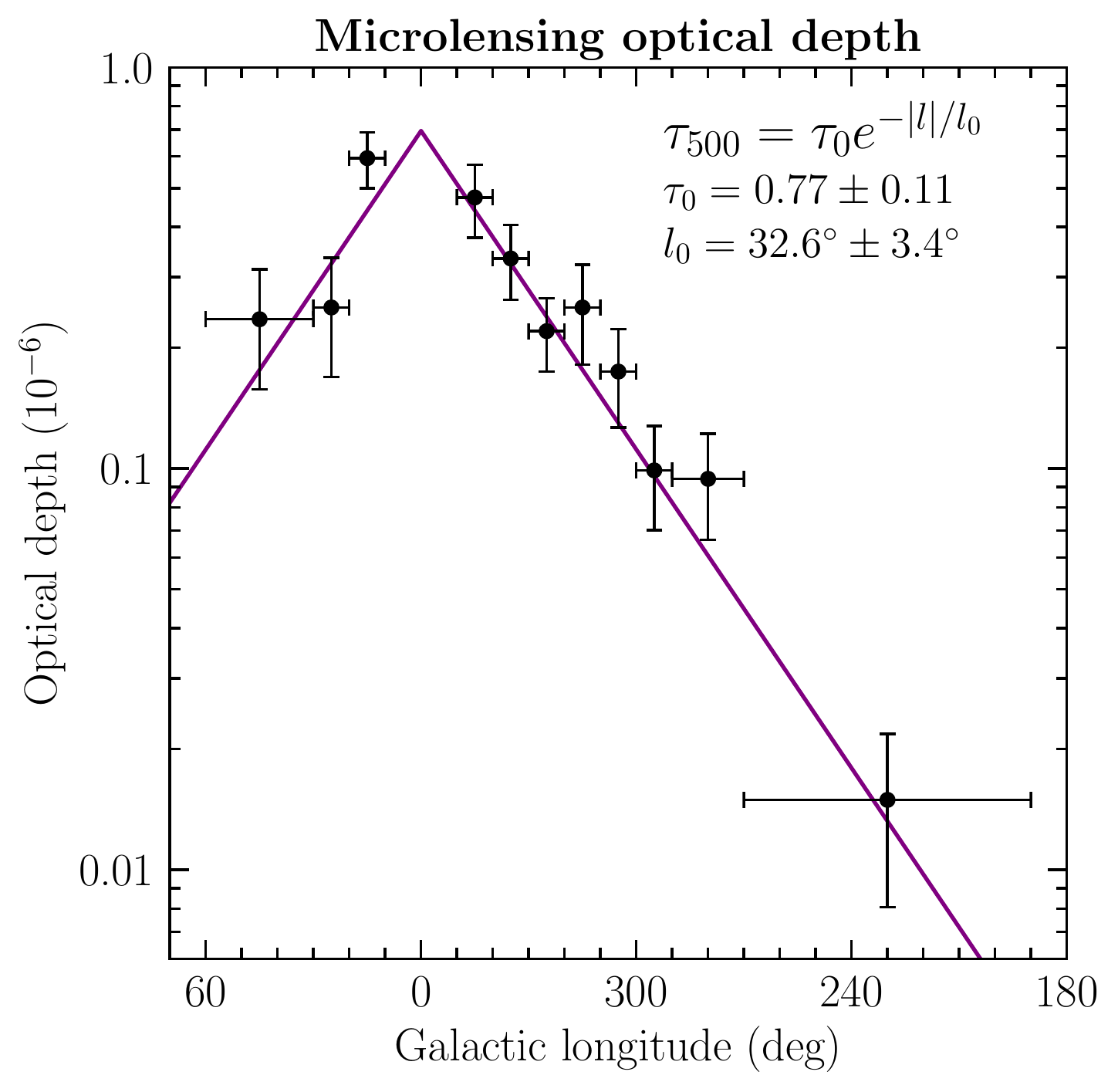}
\includegraphics[width=.49\textwidth]{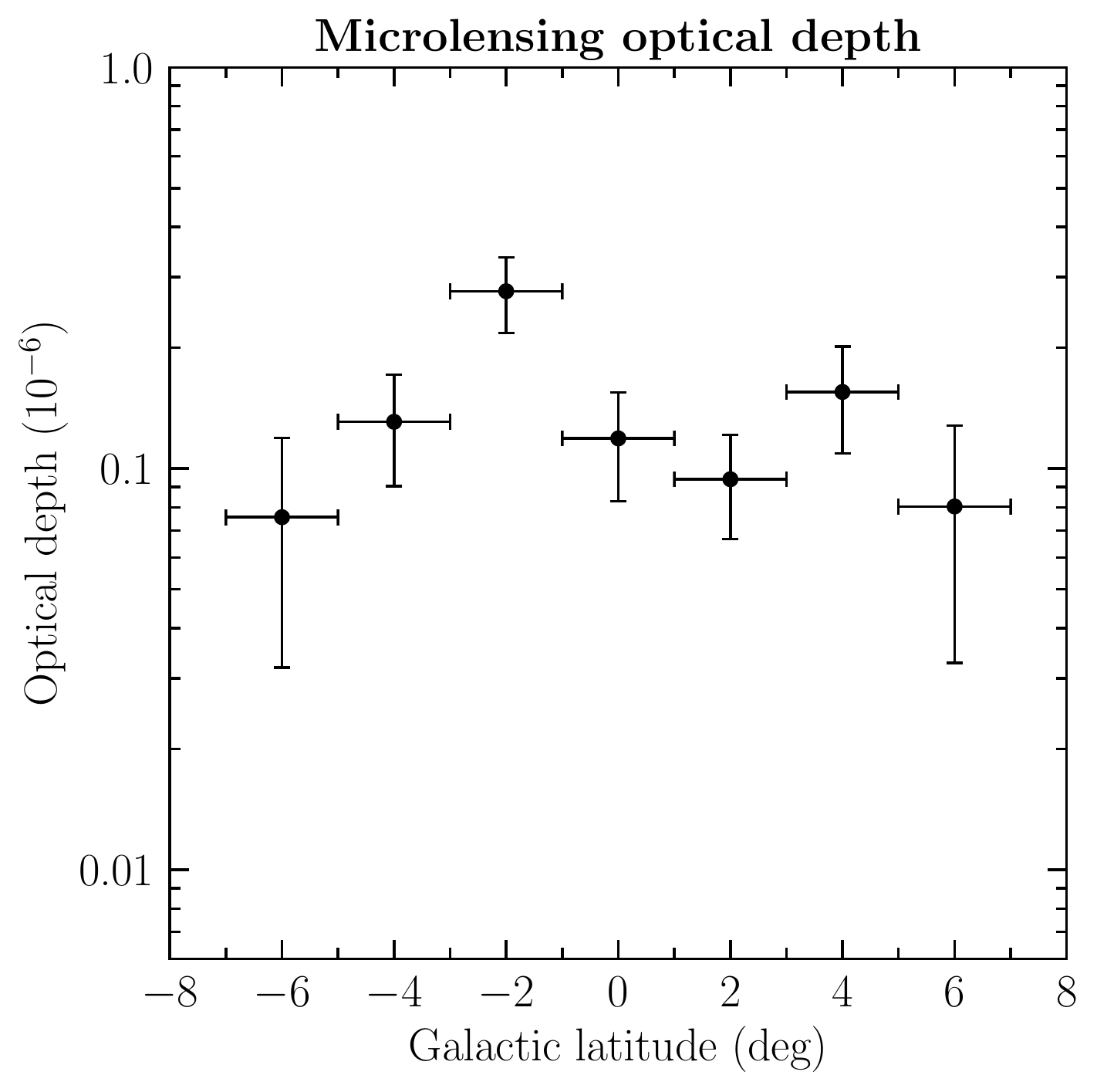}\\
\includegraphics[width=.49\textwidth]{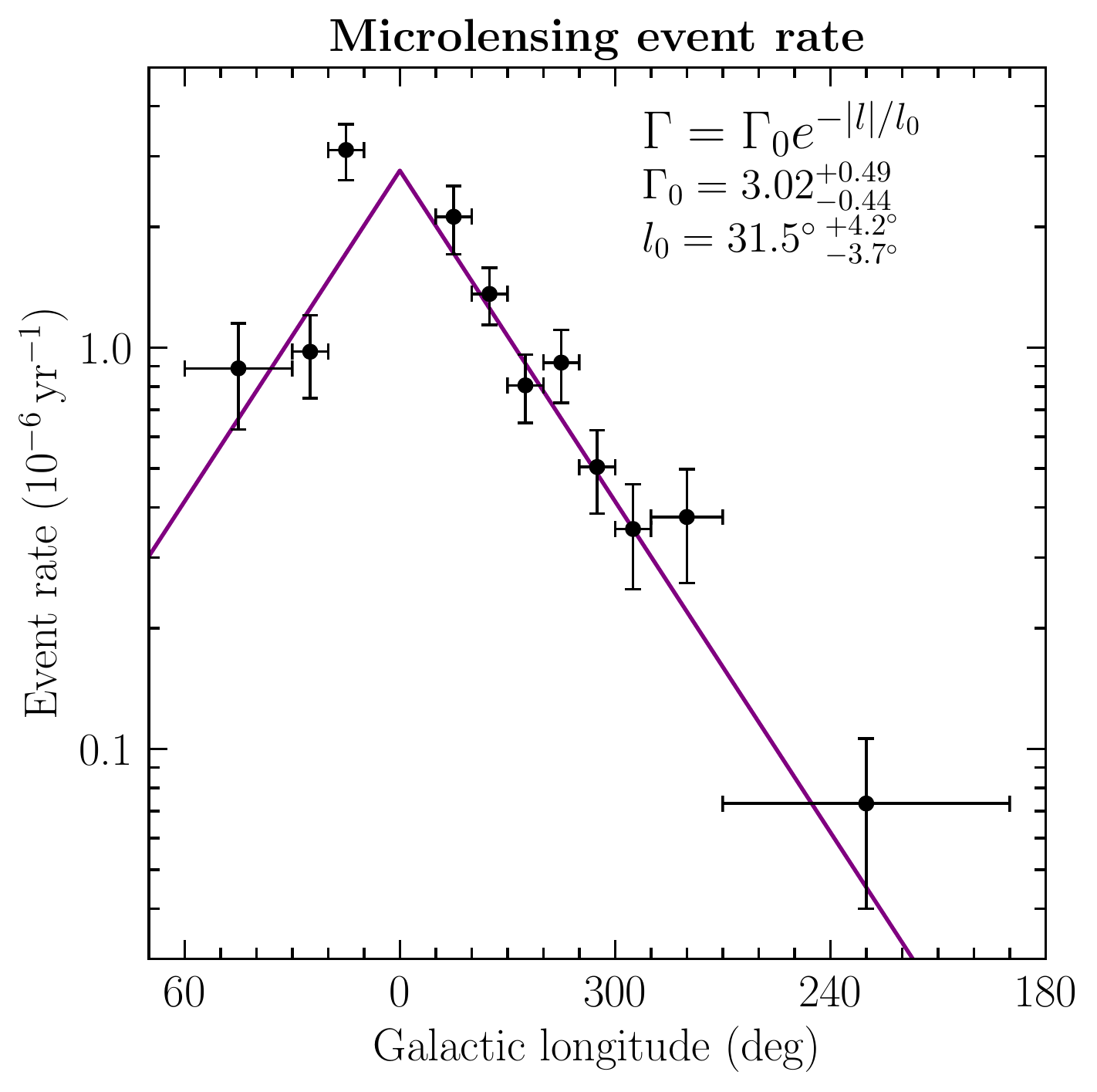}
\includegraphics[width=.49\textwidth]{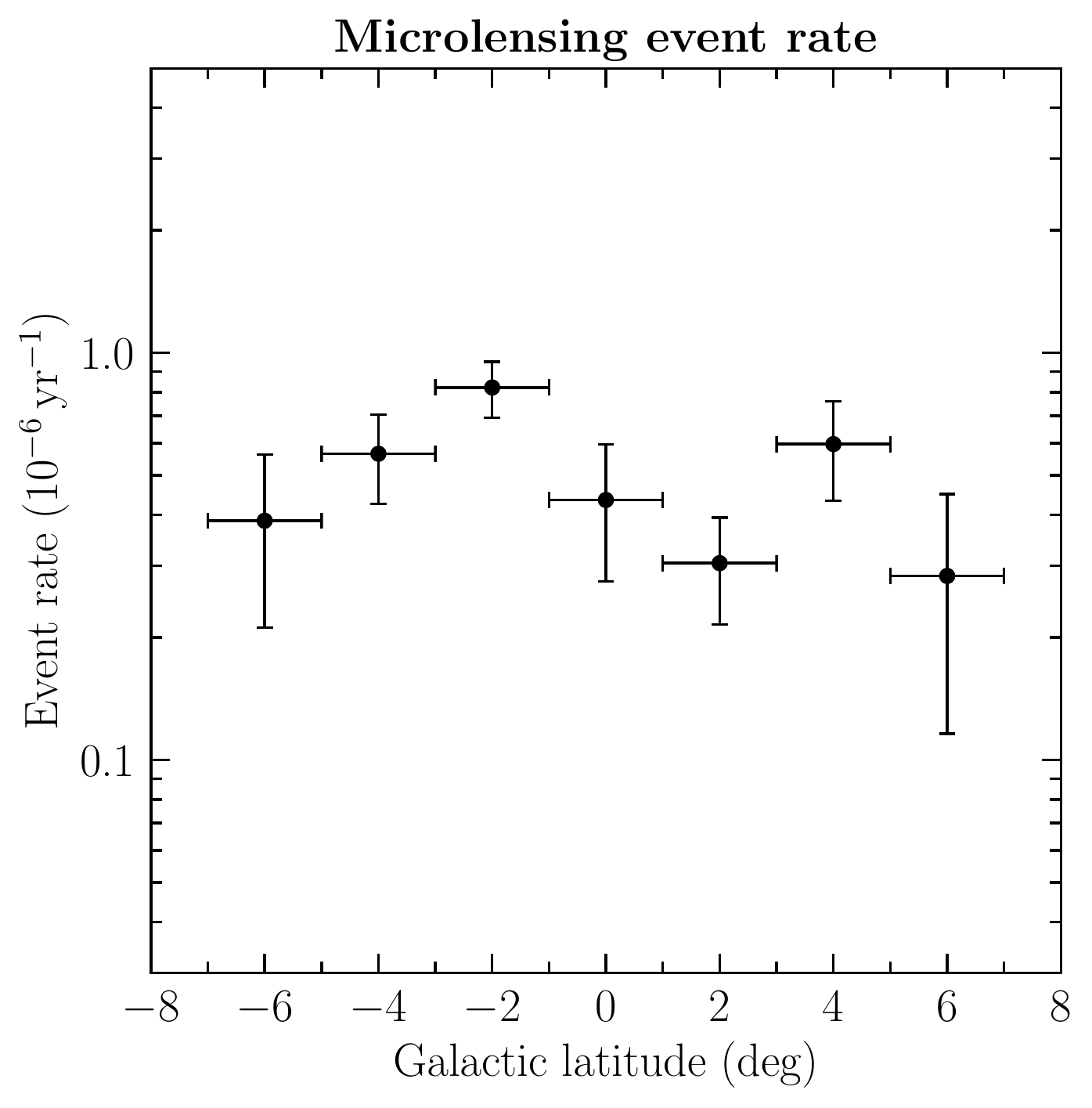}\\
\includegraphics[width=.49\textwidth]{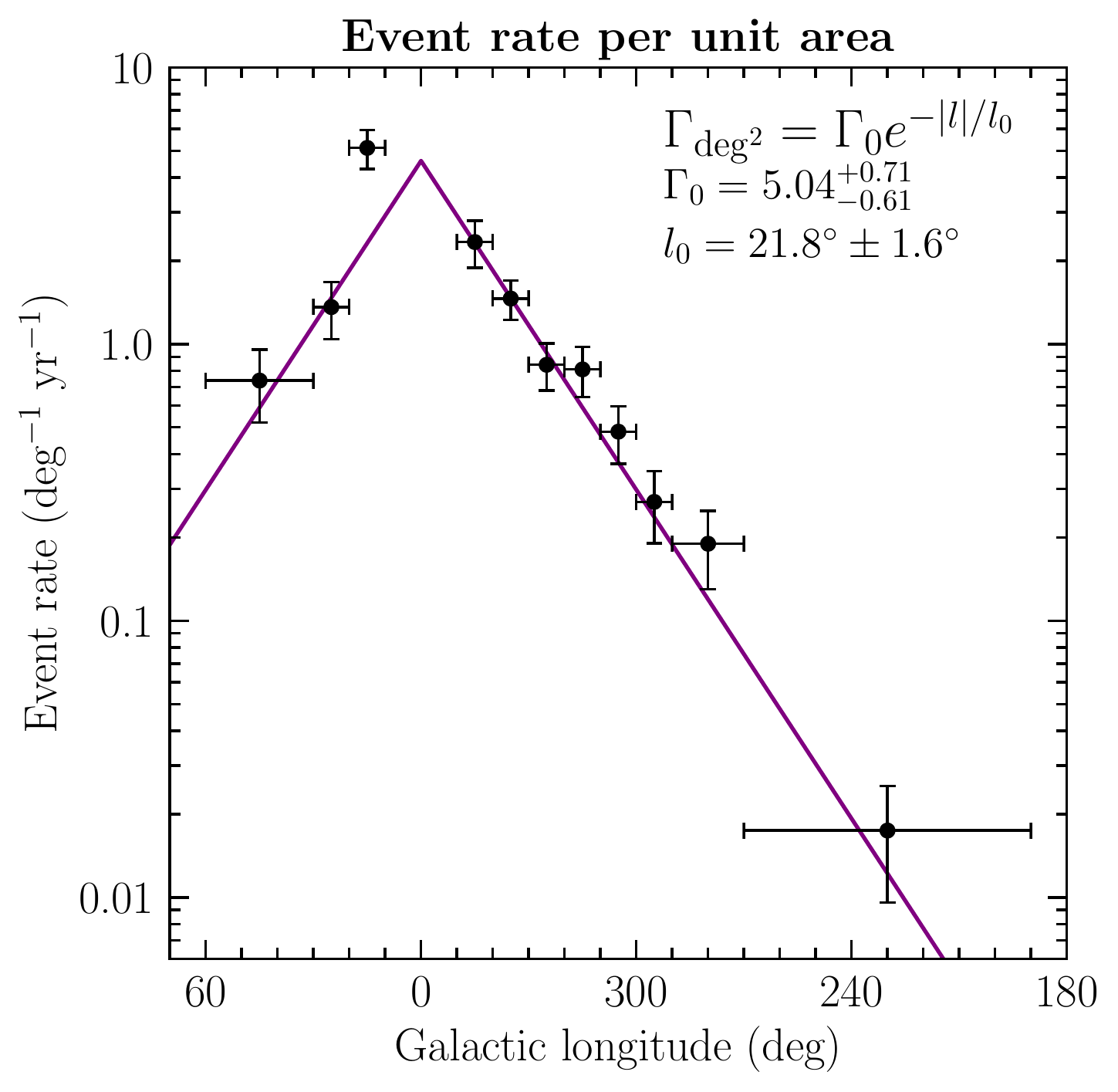}
\includegraphics[width=.49\textwidth]{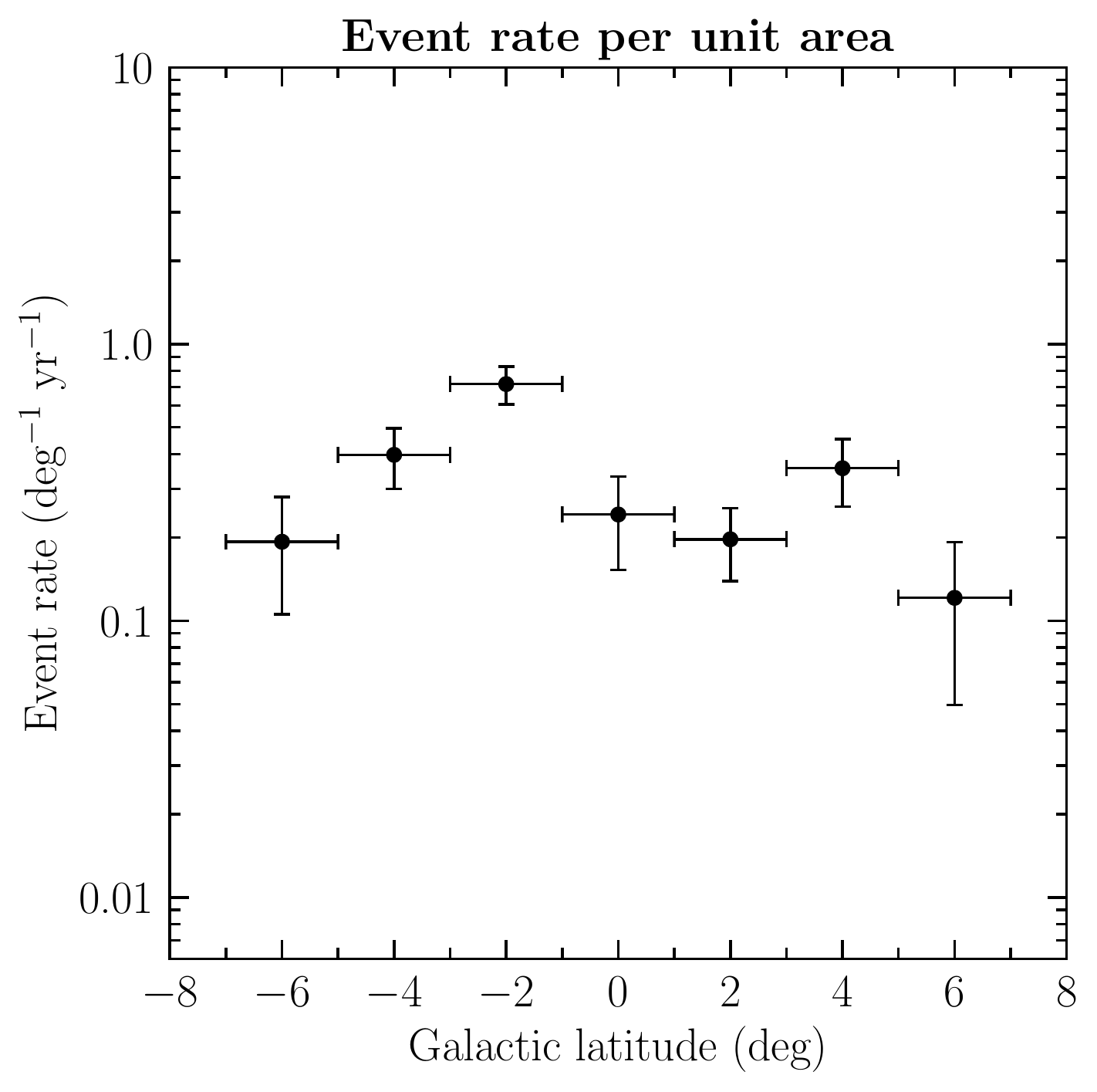}
\caption{Microlensing optical depth, event rate, and event rate per unit area as a function of Galactic longitude (left column, in the range $-7^{\circ}<b<7^{\circ}$) and latitude (right column, $240^{\circ} < l < 330^{\circ}$).}
\label{fig:tau}
\end{figure}

The measured optical depths and event rates are reported in Table~\ref{tab:tau} and plotted in the left column of Figure~\ref{fig:tau}. Each bin contains from 5 to 54 events with the median of 25 events. Owing to the modest sample size, our optical depth measurements have relatively large fractional uncertainties (from 16 to 50\% with the median of 33\%). 

Both optical depth and event rates decrease exponentially with the angular distance from the Galactic center $\tau,\Gamma\propto e^{-|l|/l_0}$. The characteristic angular scale length is $l_0=32.6^{\circ} \pm 3.4^{\circ}$ for the optical depth and $l_0=31.5^{\circ}\, ^{+4.2^{\circ}}_{-3.7^{\circ}}$ for the event rate per source. The event rate per unit area decreases faster with $l_0=21.8^{\circ} \pm 1.6^{\circ}$. Both optical depth and event rates extrapolated to $l=0^{\circ}$ ($0.77 \pm 0.11 \times 10^{-6}$, $3.02^{+0.49}_{-0.44}\times 10^{-6}$\,yr$^{-1}$, and $5.04^{+0.71}_{-0.61}$\,yr$^{-1}$\,deg$^{-2}$, respectively) are much smaller than those measured in the Galactic bulge ($\sim 2 \times 10^{-6}$, $\sim 20\times 10^{-6}$\,yr$^{-1}$, and $\sim 300$\,yr$^{-1}$\,deg$^{-2}$).

To measure the dependence of optical depth and event rates on the limiting magnitude, we chose a sample of 365~events with sources brighter than $I=20$ and recalculated their detection efficiencies (by excluding sources fainter than $I=20$ from the results of simulations described in Section \ref{sec:simuls}). Optical depths and event rates calculated using sources brighter than $I=20$ are slightly smaller than those calculated using all events: $\tau_{I \leq 20}/\tau_{I \leq 21}=\Gamma_{I \leq 20}/\Gamma_{I \leq 21}=0.86 \pm 0.13$. However, the measurements in the individual bins may vary by up to $\sim 30\%$. Because the sample size is relatively small, the optical depth and event rate per star averaged over sources brighter than $I=20$ and $I=21$ are consistent within $1\sigma$ quoted error bars.

We note that the optical depth and event rate toward $l\approx 280^{\circ}$ and $l\approx 315^{\circ}$ are slightly larger than the values predicted by our exponential model. These directions are coincident with the tangents to the Carina and Crux--Centaurus spiral arms \citep[e.g.,][]{vallee2016}, respectively, raising possibility that the increased number of microlensing events may be caused by the increased number of lenses along these lines of sight. However, the statistical significance of the optical depth excess is very small so the additional monitoring is needed to determine if that excess is real or just a statistical fluctuation. We also note that no optical depth excess is detected toward the tangent to the Norma arm ($l\approx 328^{\circ}$).

We also calculate the optical depth and event rate in seven 2$^{\circ}$ wide bins in the Galactic latitude (in the longitude range $240^{\circ} < l < 330^{\circ}$) to search for their possible variations with the distance from the Galactic equator. The results are presented in Table~\ref{tab:tau} and the right column of Figure~\ref{fig:tau}. One may expect that the optical depth should be largest at $b=0^{\circ}$, but this is not the case. The large interstellar extinction in the $I$ band limits the number of observable sources to the closest objects and so the observed optical depth at $b=0^{\circ}$ is relatively small. (A similar pattern was found by \citet{mroz2019b} in the Galactic bulge fields.) 

Both microlensing optical depth and event rate are not, however, symmetric about the Galactic equator, with $\tau(b=-2^{\circ})/\tau(b=2^{\circ})\approx 3.0$, $\Gamma(b=-2^{\circ})/\Gamma(b=2^{\circ})\approx 2.7$, and $\Gamma_{\rm deg^2}(b=-2^{\circ})/\Gamma_{\rm deg^2}(b=2^{\circ})\approx 3.6$. This asymmetry may be partially explained by the asymmetric distribution of dust \citep{marshall2006} (due to lower extinction in the southern Galactic hemisphere, we may observe sources located farther than in the northern hemisphere). However, the numbers of stars observed in both latitude bins, $N_{\rm s}(b=-2^{\circ})/N_{\rm s}(b=2^{\circ})\approx 1.4$, do not differ that much, which suggests that the asymmetric extinction cannot fully explain the observed number of microlensing events. 

We attribute the optical depth excess in the southern Galactic hemisphere to the Galactic warp. \citet{skowron2019} used distances of thousands of classical Cepheids to construct the map of the young Milky Way disk in three dimensions and they found that the majority of Cepheids in the longitude range of $240^{\circ} < l < 330^{\circ}$ lie below the Galactic equator, with the most distant Cepheids displaced by $1-1.5$\,kpc below the Galactic plane. The presence of the warp was also inferred from observations of other stellar tracers \citep[e.g.,][]{lopez2002,momany2006,reyle2009,amores2017}. The warped shape of the Galactic disk in this direction likely leads to the increased number of lenses at $b<0^{\circ}$ and so the elevated optical depth and event rates.

\subsection{Einstein timescales}

\begin{figure}
\centering
\includegraphics[width=0.48\textwidth]{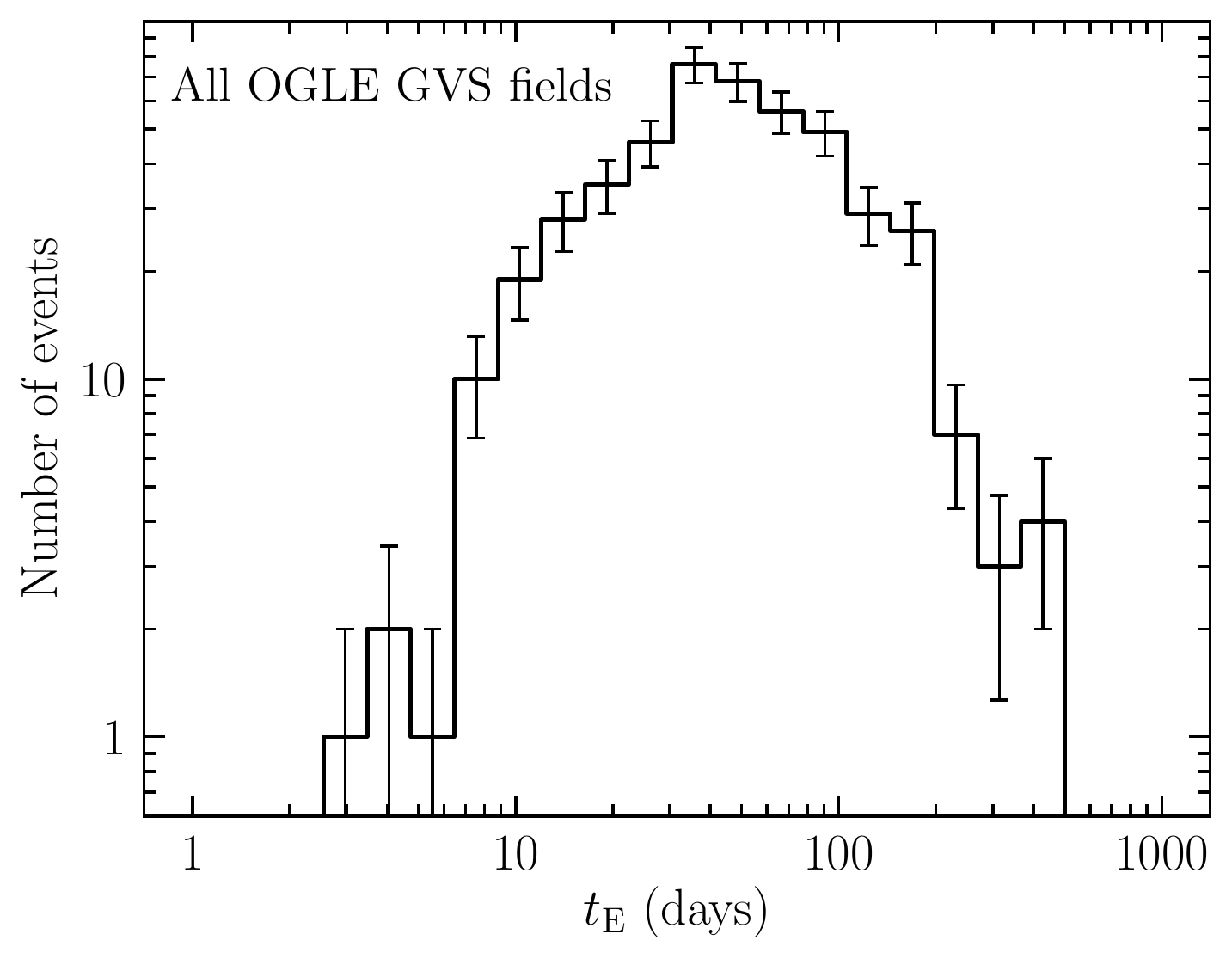}
\includegraphics[width=0.48\textwidth]{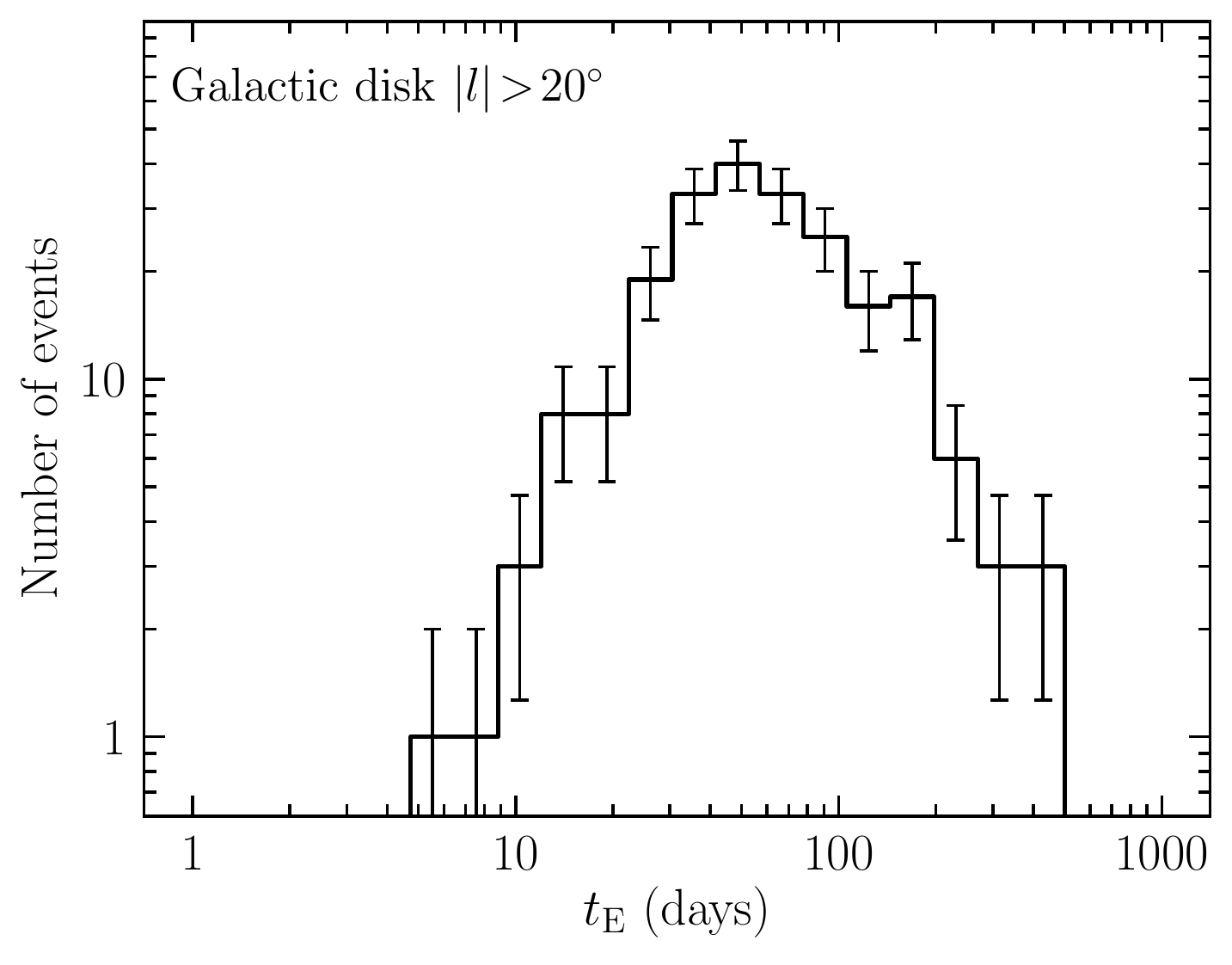}
\caption{Distribution of timescales of microlensing events detected in the OGLE GVS fields. Left panel: all 460 events. Right panel: 216 events at Galactic longitudes $|l|>20^{\circ}$.}
\label{fig:raw_timescales}
\end{figure}

\begin{figure}
\centering
\includegraphics[width=0.7\textwidth]{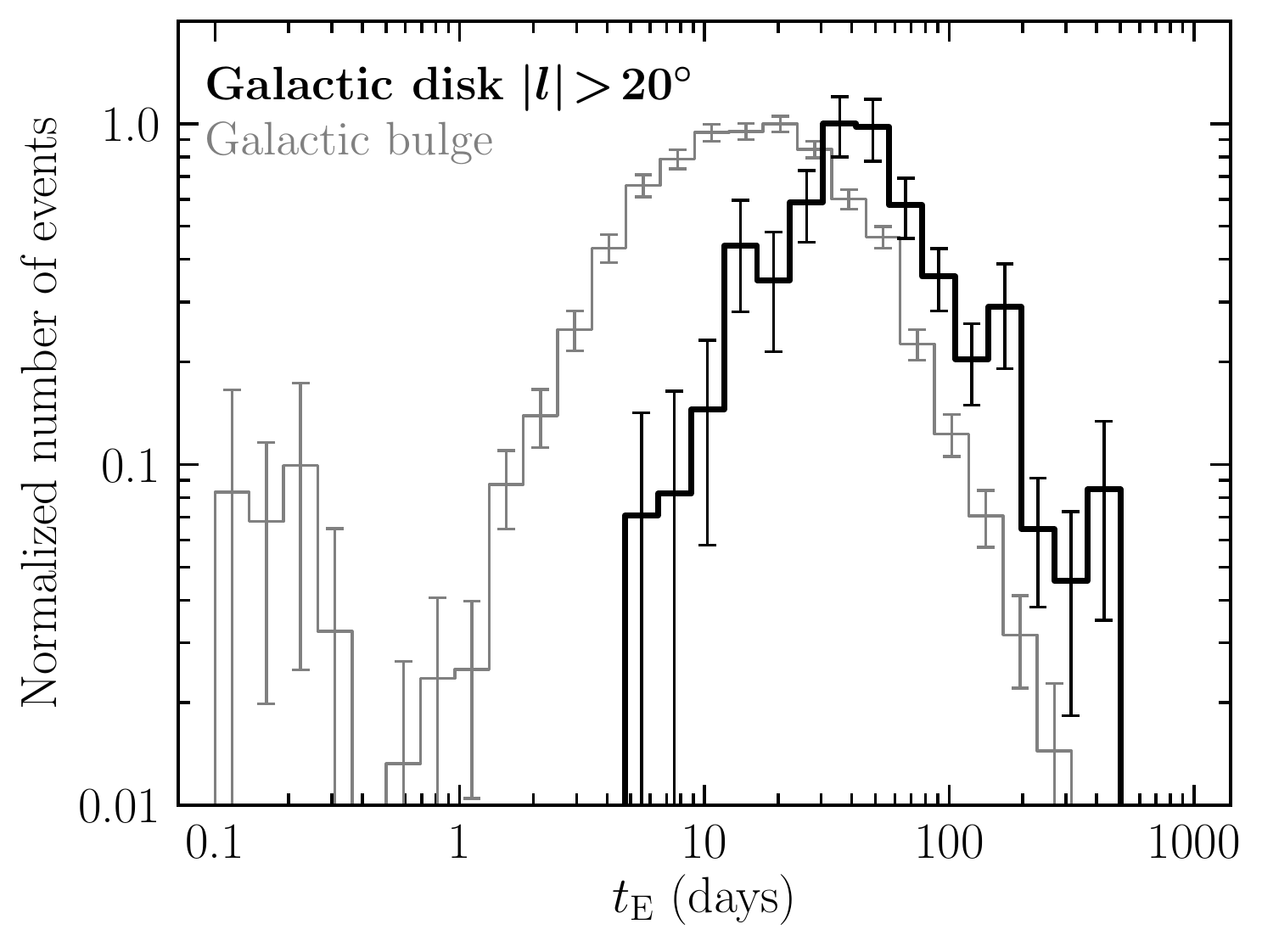}
\caption{Comparison between the detection-efficiency-corrected distributions of event timescales in the Galactic plane fields ($|l|>20^{\circ}$, thick black line, 216 events) and in the central Galactic bulge \citep[thin gray line, 2617 events;][]{mroz2017}. Galactic plane events are, on average, three times longer than those in the Galactic center.}
\label{fig:timescales}
\end{figure}

The average Einstein timescales and event timescale distribution contain useful information about the kinematics and the mass function of lenses. The distribution of the timescales of events detected in OGLE GVS fields is presented in Figure~\ref{fig:raw_timescales}. We are able to detect events as short as $\tE \approx 3$\,days and as long as $\tE \approx 500$\,days. However, the majority of detected events have timescales in the range of $10<\tE<200$\,days. Figure~\ref{fig:timescales} presents the detection-efficiency-corrected distributions of event timescales in the Galactic plane fields ($|l|>20^{\circ}$) and in the central Galactic bulge \citep{mroz2017}. Both histograms have a similar shape (slopes of short- and long-timescale tails), but events in the Galactic disk are longer. The average Einstein timescale in the Galactic plane fields, defined as 
\begin{equation}
\langle\tE\rangle=\frac{\sum_i t_{{\rm E},i}/\varepsilon(t_{{\rm E},i})}{\sum_i 1/\varepsilon(t_{{\rm E},i})},
\end{equation}
is equal to $61.5 \pm 5.0$\,day and is approximately three times longer than that in the Galactic bulge. In particular, our sample contains only two events with $\tE<10$\,day at $|l|>20^{\circ}$, with timescales about 5.7 and 7.2\,days (see right panel of Figure~\ref{fig:raw_timescales}). Our detection efficiency in the disk fields at these timescales is small but nonnegligible, which is demonstrated by the detection of 19 events with $\tE<10$\,day in the outer Galactic bulge fields (observed with a cadence similar to that in the Galactic plane; the shortest-timescale event has $\tE \approx 3.3$\,days, as shown in the left panel of Figure~\ref{fig:raw_timescales}). This apparent lack of short-timescale events is a distinct feature of Galactic disk microlensing events.

\citet{mroz2019b} measured that the mean timescales of microlensing events in the Galactic bulge increase with the increasing Galactic longitude: from $\sim 23$\,days at $l=0^{\circ}$ to $\sim 36$\,days at $l=10^{\circ}$. The average timescales of microlensing events in the analyzed Galactic longitude bins (Table~\ref{tab:tau}) follow that trend. The mean timescales in the two bins located nearest the Galactic center ($340^{\circ}<l<350^{\circ}$ and $10^{\circ}<l<20^{\circ}$) are longer: $52.0 \pm 7.8$\,days and $44.4 \pm 4.6$\,days, respectively. At larger Galactic longitudes ($|l|>20^{\circ}$), the average Einstein timescales reach a value of $\langle\tE\rangle=61.5 \pm 5.0$\,days with little dependence on the Galactic longitude.
This is further demonstrated by the facts that
\begin{equation}
\langle\tE\rangle=\frac{2\tau}{\pi\Gamma}
\end{equation}
and that both $\tau$ and $\Gamma$ exponentially decrease with the distance from the Galactic center with the approximately the same angular scale length of $\sim 32^{\circ}$.

\subsection{Comparison with previous measurements and theoretical models}

\citet{rahal2009} published estimates of the microlensing optical depth toward seven sight lines in the Galactic disk: $l=307^{\circ}$ ($\theta$ Mus), $l=331^{\circ}$ ($\gamma$ Nor), $l=19^{\circ}$ ($\gamma$ Sct), and $l=27^{\circ}$ ($\beta$ Sct) based on observations collected by the EROS survey (see Table~\ref{tab:eros}). Their measurements were based on a small number of 3--10 events. EROS observed 29 Galactic disk fields using their custom $B_{\rm EROS}$ and $R_{\rm EROS}$ filters with longer exposure times ($120-180$\,s) than those used by the OGLE GVS, which enabled them to detect sources as faint as $R_{\rm EROS} \approx I \approx 22$.

We measured the optical depth and average Einstein timescales in regions corresponding to the EROS fields (see Table~\ref{tab:eros}.) Our measurements are consistent with those of \citet{rahal2009} within the quoted error bars, which are relatively large given the small statistics (OGLE measurements differ by $-0.9\sigma$, $-1.0\sigma$, $-1.2\sigma$, and $+0.8\sigma$). These differences can be usually tracked down to a single long-timescale event, which contributes to a large fraction of the measured optical depth. For example, \citet{rahal2009} found three events in the $\theta$\,Mus region, one of which has $\tE \approx 205$\,days and contributes to 64\% of the total optical depth measured in that region. 

\begin{deluxetable}{lrrrrrrrr}[h]
\tablecaption{Microlensing Optical Depth and Average Timescales in EROS Fields \citep{rahal2009} \label{tab:eros}}
\tablehead{
\colhead{Region} & \colhead{$l$} & \colhead{$b$} & \colhead{$\tau_{\rm EROS}$} & \colhead{$\langle\tE\rangle_{\rm EROS}$} & \colhead{$N_{\rm EROS}$}  & \colhead{$\tau_{\rm OGLE}$} & \colhead{$\langle\tE\rangle_{\rm OGLE}$} & \colhead{$N_{\rm OGLE}$} \\ 
\colhead{} & \colhead{(deg)} & \colhead{(deg)} & \colhead{($10^{-6}$)} & \colhead{(days)} & \colhead{} & \colhead{($10^{-6}$)} & \colhead{(days)} & \colhead{}
}
\startdata
$\theta$ Mus & 306.56 & $-1.46$ & $0.67 ^{+0.63}_{-0.52}$ & $97 \pm 47$ & 3  & $0.20 \pm 0.08$ & $55 \pm  8$ &  7 \\
$\gamma$ Nor & 331.09 & $-2.42$ & $0.49 ^{+0.21}_{-0.18}$ & $57 \pm 10$ & 10 & $0.28 \pm 0.09$ & $60 \pm 11$ & 12 \\
$\gamma$ Sct &  18.51 & $-2.09$ & $0.72 ^{+0.41}_{-0.28}$ & $47 \pm  6$ & 6  & $0.32 \pm 0.19$ & $71 \pm 22$ &  3 \\
$\beta$ Sct  &  26.60 & $-2.15$ & $0.30 ^{+0.23}_{-0.20}$ & $59 \pm  6$ & 3  & $0.61 \pm 0.32$ & $87 \pm 44$ & 11 \\
\enddata
\end{deluxetable}

\citet{sajadian2019} published theoretical predictions for the detection and characterization of microlensing events in the Galactic disk by the Rubin Observatory. They simulated an ensemble of events toward over 64,000 directions in the Galactic plane ($|l|\leq 100^{\circ}$, $|b|\leq 10^{\circ}$), generated their synthetic light curves, and studied their detectability under different proposed observing strategies. 

The Rubin Observatory is expected to detect much fainter events ($r\approx 24.3$ in the baseline) than those observed by OGLE GVS; filters and pixel size are also different. With that in mind, we used simulations of \citet{sajadian2019} to compute the average optical depth in the range of $|b|\leq 7^{\circ}$ as a function of Galactic longitude by calculating the weighed mean of individual grid points (we chose the predicted $r$-band star counts as weights). The average optical depth decreases exponentially at $|l|>20^{\circ}$ with the characteristic angular scale length of $35.8^{\circ}$, which is quite consistent with the value we measured using OGLE GVS data ($32.6^{\circ} \pm 3.4^{\circ}$; Figure~\ref{fig:tau}). Similarly, the theoretical event rate per star and event rate per unit area decrease exponentially with angular scale lengths of $33.1^{\circ}$ and $16.7^{\circ}$, which are similar to the values we measured ($31.5^{\circ}\,^{+4.2^{\circ}}_{-3.7^{\circ}}$ and $21.8^{\circ} \pm 1.6^{\circ}$, respectively). However, the normalizations of both theoretical optical depth and event rates do not match our observations. This is expected because the deeper observations by the Rubin Observatory enable probing a larger volume of the Galaxy.

\citet{sajadian2019} also found that average Einstein timescales of microlensing events detectable by the Rubin Observatory should gradually increase from $\sim 27$\,days at $l=0^{\circ}$ to $\sim 70$\,days at $l=90^{\circ}$. This trend does not match our observations, in which we find that mean timescales reach $\tE\approx 60$\,days at $|l|\approx 30^{\circ}$. However, \citet{sajadian2019} report the average Einstein timescales of detectable microlensing events (uncorrected for selection criteria) by assuming a fiducial survey cadence of 3\,days, so it is unclear how to compare these numbers with our observations. Another possible explanation is that sources observable by the Rubin Observatory are on average farther away than those detected in the OGLE GVS, resulting in a different proper motion distribution (and thus a different average timescale).

\section{Summary and conclusions}

Gravitational microlensing surveys have been traditionally observing the central regions of the Galactic bulge, where the event rate is the highest. It was hypothesized that many microlensing events should occur in the Galactic plane far from the Galactic center, but their detection was deemed challenging, mostly because of practical considerations. Finding microlensing events requires frequent monitoring of a large area along the Galactic plane. This is why the first Galactic plane events were detected mostly serendipitously \citep[e.g.,][]{fukui2007,gaudi_halloween2008, nucita2018,fukui2019,zang2019}. Only the recent few years have brought about more detections of microlensing events in the Galactic plane, mostly thanks to efforts by the \textit{Gaia} and ZTF groups.

In 2013, the OGLE collaboration has initiated the Galaxy Variability Survey (GVS) -- a survey dedicated for the study of the variability of stars located in the Galactic plane ($|b|<7^{\circ}$, $0^{\circ}<l<50^{\circ}$, $190^{\circ}<l<360^{\circ}$) and in an extended area around the outer Galactic bulge. Thus far, the survey led to the discovery of thousands of new variable stars -- for example Cepheids \citep{udalski2018cep,skowron2019} and RR Lyrae stars \citep{soszynski2019}. The majority of GVS fields were observed 100--200 times during a period of 2--7 yr, rendering it possible for us to search for microlensing events (and to distinguish them from other astrophysical sources). 

In this paper, we have presented the results of the first comprehensive search for Galactic plane microlensing events in an area of almost 3000 square degrees. We have found 460 microlensing events fulfilling objective selection criteria and additional 170 possible events that were identified by the visual inspection of their light curves. All light curves, in addition to star counts, detection efficiencies, and measured microlensing statistics, are publicly available to the astronomical community at 
\begin{center}
\url{http://www.astrouw.edu.pl/ogle/ogle4/galactic_disk_microlensing}.
\end{center}
We run extensive catalog-level simulations of detectability of microlensing events in our experiment (over $\sim 3$ billion light curves were simulated), which enabled us to study the global properties of microlensing events in the Galactic plane for the first time.

We demonstrate that the average Einstein timescales of Galactic plane microlensing events are on average three times longer than those of Galactic bulge events, with little dependence on the Galactic longitude. This property was expected from the theoretical point of view because lensing objects are, typically, closer than those toward the Galactic bulge (and so their Einstein radii are larger). Moreover, as an observer, lens, and source -- all located in the Galactic disk -- are moving in a similar direction, the relative lens-source proper motions should be lower than those in the Galactic bulge.

This has several interesting consequences, some of which were previously discussed \citep[e.g.,][]{gould_lsst2013}. Longer timescales (the average event timescale in the Galactic disk is $\sim 61.5$\,days, compared to $\sim 20$\,days in the central bulge) facilitate the measurement of the annual microlens parallax effect (and so the mass and distance to the lens). (The analysis of parallax effects in the sample of microlensing events described in this paper will be published in a separate study.) Galactic disk events may be followed up with a lower cadence but the time needed for the lens and source to separate is normally longer than in the Galactic bulge. \citet{mroz2017} found a few ultra-short-timescale microlensing events ($\tE=0.1-0.3$\,days), which may be attributed to free-floating or wide-orbit planets. Similar events, observed in the Galactic plane, should also be longer. While the sensitivity of the OGLE GVS survey to such short-timescale events is nearly zero, some wide-field surveys (for example ZTF) are observing the Galactic plane with a higher cadence.

Another interesting aspect of Galactic plane microlensing is the larger angular Einstein radii of lenses compared to Galactic bulge events, making it easier to detect the astrometric microlensing signal. All objects reported in this paper should have been concurrently observed by \textit{Gaia}. Combining OGLE light curves and \textit{Gaia} data should enable one to measure Einstein radii, and therefore masses, for a significant fraction of reported events \citep[24/71 events are brighter than $I=16/17$ in the baseline;][]{rybicki2018}.

We also measure the microlensing optical depth and event rate as a function of Galactic longitude and demonstrate that they exponentially decrease with the angular distance from the Galactic Center (with the characteristic angular scale length of $\sim 32^{\circ}$). This is in good agreement with the expectations of the Galactic models \citep{sajadian2019}. The average optical depth decreases from $0.5\times 10^{-6}$ at $l=10^{\circ}$ to $1.5\times 10^{-8}$ in the Galactic anticenter. We also find that the optical depth in the longitude range of $240^{\circ}<l<330^{\circ}$ is asymmetric about the Galactic equator, which we interpret as a signature of the Galactic warp. Finally, we also find a small optical depth excess toward $l\approx 280^{\circ}$ and $l\approx 315^{\circ}$ -- that is, directions tangent to the Carina and Crux--Centaurus spiral arms. However, the statistical significance of that excess is small, so more observations are needed to confirm this finding.

Our measurements can be extended by other, current, and planned surveys. ZTF is currently conducting high-cadence observations of the northern Galactic plane, which would complement southern-hemisphere-based OGLE data. The Galactic plane is also observed by \textit{Gaia}, albeit with a very low cadence. Finally, it was proposed to observe the southern Galactic plane with the Rubin Observatory \citep[e.g.,][]{gould_lsst2013,street2018}. Results presented in this paper may inform the planning of such surveys.

\section*{Acknowledgements}

We thank the anonymous referee for his/her comments, which improved the presentation of our results. We thank M.~Kubiak, G.~Pietrzy\'nski, \L{}.~Wyrzykowski, and M.~Pawlak for their contribution to the collection of the OGLE photometric data analyzed in this paper. P.M. acknowledges support from the National Science Center, Poland (grant ETIUDA 2018/28/T/ST9/00096). The OGLE project has received funding from the National Science Center, Poland, grant MAESTRO 2014/14/A/ST9/00121 to A.U. 

\bibliographystyle{aasjournal}
\bibliography{sample}

\appendix
\restartappendixnumbering 

\section{OGLE GVS Fields}

\begin{deluxetable}{lrrrrrr}[h]
\tablecaption{Basic Information about Analyzed Fields \label{tab:allfields}}
\tablehead{
\colhead{Field} & \colhead{R.A.} & \colhead{Decl.} & \colhead{$l$} &
\colhead{$b$} & \colhead{$N_{\mathrm{stars}}$} & \colhead{$N_{\mathrm{epochs}}$} \\
\colhead{} & \colhead{(J2000, deg)} & \colhead{(J2000, deg)} & \colhead{(deg)} &
\colhead{(deg)} & \colhead{($10^6$)} & \colhead{}
}
\startdata
DG1000 & 279.76667 & --3.42222 & 28.47547 & 1.18686 & 1.08 & 128 \\
DG1001 & 279.77500 & --4.65278 & 27.38541 & 0.61553 & 1.45 & 128 \\
DG1002 & 279.78333 & --5.88333 & 26.29558 & 0.04401 & 0.83 & 126 \\
DG1003 & 279.79167 & --7.11389 & 25.20574 & --0.52749 & 0.97 & 121 \\
DG1004 & 279.80000 & --8.34444 & 24.11568 & --1.09877 & 1.30 & 126 \\
DG1005 & 279.80833 & --9.57500 & 23.02518 & --1.66960 & 1.60 & 128 \\
DG1006 & 279.81667 & --10.80556 & 21.93402 & --2.23977 & 2.05 & 128 \\
DG1007 & 280.97917 & --1.57639 & 30.67132 & 0.95385 & 0.81 & 124 \\
DG1008 & 280.98750 & --2.80694 & 29.58037 & 0.38434 & 0.70 & 127 \\
DG1009 & 281.00000 & --4.03750 & 28.49146 & --0.18899 & 0.92 & 126 \\
DG1010 & 281.01250 & --5.26806 & 27.40246 & --0.76222 & 2.16 & 126 \\
DG1011 & 281.02083 & --6.49861 & 26.31126 & --1.33145 & 2.97 & 128 \\
DG1012 & 281.03333 & --7.72917 & 25.22141 & --1.90383 & 2.42 & 127 \\
DG1013 & 281.04583 & --8.95972 & 24.13082 & --2.47547 & 2.57 & 125 \\
DG1014 & 281.05833 & --10.19028 & 23.03924 & --3.04612 & 1.80 & 126 \\
DG1015 & 282.20833 & --0.96111 & 31.77964 & 0.14119 & 0.87 & 129 \\
DG1016 & 282.21667 & --2.19167 & 30.68831 & --0.42746 & 0.28 & 128 \\
DG1017 & 282.22917 & --3.42222 & 29.59871 & --0.99964 & 0.43 & 127 \\
DG1018 & 282.23750 & --4.65278 & 28.50683 & --1.56774 & 1.22 & 130 \\
DG1019 & 282.25000 & --5.88333 & 27.41622 & --2.13893 & 2.51 & 126 \\
\dots & \dots & \dots & \dots & \dots & \dots & \dots \\
\enddata
\tablecomments{Equatorial coordinates are given for the epoch J2000. Here $N_{\rm stars}$ is the number of stars in the database in millions, and $N_{\rm epochs}$ is the number of collected frames used in the analysis; $l$ and $b$ are Galactic longitude and latitude, respectively. (This table is available in its entirety in machine-readable form.)}
\end{deluxetable}

\clearpage

\section{Microlensing events in the OGLE GVS fields}
\restartappendixnumbering 

\begin{deluxetable}{@{}lrrrrrrrl@{}}[h]
\tablecaption{Best-fitting Parameters of the Analyzed Microlensing Events in the OGLE GVS Fields. \label{tab:params}}
\tabletypesize{\scriptsize}
\tablehead{
\colhead{Star} & \colhead{R.A.} & \colhead{Decl.} & \colhead{$t_0$ (HJD)} & \colhead{$\tE$ (days)} & \colhead{$u_0$} & \colhead{$I_{\rm s}$} & \colhead{$f_{\rm s}$} & \colhead{IDs}
}
\startdata
GD1793.08.3677 & \ra{06}{37}{40}{01} & \dec{+13}{57}{18}{5} & $2457689.76^{+3.47}_{-3.50}$ & $69.81^{+8.36}_{-5.23}$ & $0.314^{+0.047}_{-0.056}$ & $17.18^{+0.36}_{-0.21}$ & $0.77^{+0.17}_{-0.22}$ & ASASSN-16li \\
GD1705.29.3640 & \ra{06}{59}{57}{53} & \dec{-04}{27}{44}{8} & $2457740.21^{+0.36}_{-0.35}$ & $26.80^{+6.75}_{-4.25}$ & $0.336^{+0.106}_{-0.096}$ & $18.43^{+0.46}_{-0.41}$ & $0.58^{+0.26}_{-0.20}$ & - \\
GD1638.20.135 & \ra{07}{29}{46}{15} & \dec{-19}{40}{06}{8} & $2458454.40^{+10.48}_{-12.71}$ & $86.21^{+12.71}_{-8.66}$ & $0.604^{+0.121}_{-0.142}$ & $15.54^{+0.55}_{-0.31}$ & $0.67^{+0.23}_{-0.27}$ & - \\
GD1533.19.13800 & \ra{08}{18}{08}{32} & \dec{-28}{41}{50}{5} & $2456405.75^{+1.48}_{-1.58}$ & $39.43^{+10.16}_{-6.34}$ & $0.306^{+0.064}_{-0.077}$ & $20.25^{+0.39}_{-0.27}$ & $1.06^{+0.30}_{-0.32}$ & - \\
GD1486.09.10016 & \ra{08}{52}{28}{09} & \dec{-48}{45}{24}{9} & $2457751.21^{+1.47}_{-1.84}$ & $47.35^{+18.37}_{-9.31}$ & $0.288^{+0.080}_{-0.107}$ & $20.19^{+0.60}_{-0.34}$ & $0.87^{+0.32}_{-0.37}$ & - \\
GD1454.01.15973 & \ra{09}{17}{57}{59} & \dec{-54}{00}{49}{4} & $2456454.99^{+6.34}_{-4.18}$ & $49.86^{+19.64}_{-10.26}$ & $0.686^{+0.219}_{-0.291}$ & $18.90^{+0.95}_{-0.56}$ & $0.52^{+0.35}_{-0.30}$ & - \\
GD1446.16.5034 & \ra{09}{18}{06}{73} & \dec{-54}{26}{56}{0} & $2458553.05^{+0.15}_{-0.16}$ & $75.51^{+11.36}_{-8.81}$ & $0.076^{+0.013}_{-0.012}$ & $19.03^{+0.19}_{-0.17}$ & $0.79^{+0.14}_{-0.13}$ & Gaia19bek \\
GD1446.23.3493 & \ra{09}{19}{53}{98} & \dec{-54}{13}{14}{3} & $2458575.32^{+0.43}_{-0.41}$ & $42.02^{+9.97}_{-5.46}$ & $0.202^{+0.051}_{-0.054}$ & $18.84^{+0.37}_{-0.28}$ & $0.74^{+0.22}_{-0.21}$ & Gaia19bej \\
GD1445.04.2394 & \ra{09}{23}{33}{76} & \dec{-53}{32}{42}{7} & $2456744.64^{+2.76}_{-2.71}$ & $81.72^{+23.96}_{-12.39}$ & $0.569^{+0.126}_{-0.179}$ & $18.93^{+0.61}_{-0.36}$ & $0.68^{+0.27}_{-0.29}$ & - \\
GD2073.13.21817 & \ra{09}{27}{55}{80} & \dec{-60}{24}{16}{8} & $2457851.93^{+0.04}_{-0.05}$ & $5.99^{+2.44}_{-1.24}$ & $0.049^{+0.028}_{-0.025}$ & $20.75^{+0.47}_{-0.33}$ & $1.12^{+0.40}_{-0.39}$ & - \\
GD1431.20.13279 & \ra{09}{36}{48}{11} & \dec{-56}{30}{44}{1} & $2456736.49^{+0.46}_{-0.49}$ & $31.66^{+5.97}_{-4.01}$ & $0.052^{+0.014}_{-0.015}$ & $20.10^{+0.27}_{-0.22}$ & $1.22^{+0.26}_{-0.27}$ & - \\
GD1430.03.6048 & \ra{09}{37}{25}{19} & \dec{-56}{04}{53}{5} & $2456729.24^{+2.61}_{-2.52}$ & $87.54^{+68.05}_{-20.04}$ & $0.253^{+0.087}_{-0.128}$ & $20.60^{+0.91}_{-0.42}$ & $1.09^{+0.51}_{-0.61}$ & - \\
GD1408.08.10041 & \ra{09}{55}{52}{64} & \dec{-56}{49}{34}{3} & $2456727.03^{+0.04}_{-0.04}$ & $47.40^{+6.23}_{-5.18}$ & $0.041^{+0.006}_{-0.006}$ & $19.89^{+0.16}_{-0.15}$ & $0.28^{+0.04}_{-0.04}$ & - \\
GD2054.14.1066 & \ra{10}{17}{59}{12} & \dec{-61}{50}{15}{9} & $2458151.83^{+0.03}_{-0.03}$ & $91.92^{+10.11}_{-9.03}$ & $0.032^{+0.004}_{-0.004}$ & $19.69^{+0.13}_{-0.13}$ & $0.20^{+0.02}_{-0.02}$ & - \\
GD1387.31.13599 & \ra{10}{21}{14}{63} & \dec{-58}{37}{38}{2} & $2457602.65^{+12.38}_{-11.86}$ & $409.43^{+112.17}_{-55.59}$ & $0.752^{+0.136}_{-0.215}$ & $18.55^{+0.60}_{-0.33}$ & $0.69^{+0.25}_{-0.29}$ & - \\
\dots & \dots & \dots & \dots & \dots & \dots & \dots & \dots  & \dots \\
\enddata
\tablecomments{For each parameter, we provide the median and $1\sigma$ confidence interval derived from the marginalized posterior distribution from the Monte Carlo chain. Here $I_{\rm s}$ is the source brightness and $f_{\rm s}=\Fs/(\Fs+\Fb)$ is the blending parameter. Equatorial coordinates are given for the epoch J2000. (This table is available in its entirety in machine-readable form.)}
\end{deluxetable}

\begin{deluxetable}{@{}lrrl@{}}[h]
\tablecaption{Possible Microlensing Events. \label{tab:possible}}
\tabletypesize{\scriptsize}
\tablehead{
\colhead{Star} & \colhead{R.A.} & \colhead{Decl.} & \colhead{Remarks}
}
\startdata
GD1823.31.4508 & \ra{06}{17}{55}{58} & \dec{+20}{13}{07}{9} & binary \\
GD1648.06.5392 & \ra{07}{22}{58}{82} & \dec{-15}{51}{39}{4} & - \\
GD1606.28.7724 & \ra{07}{44}{22}{37} & \dec{-28}{26}{36}{3} & Gaia17aqu \\
GD1537.21.175 & \ra{08}{16}{11}{24} & \dec{-33}{50}{35}{0} & - \\
GD2110.04.10077 & \ra{08}{29}{51}{45} & \dec{-47}{11}{32}{0} & incomplete light curve \\
GD1454.26.351 & \ra{09}{17}{47}{11} & \dec{-53}{17}{06}{5} & - \\
GD2057.10.16399 & \ra{10}{12}{07}{51} & \dec{-62}{14}{35}{0} & - \\
GD1368.18.10432 & \ra{10}{53}{46}{91} & \dec{-59}{30}{35}{0} & - \\
GD1353.23.15634 & \ra{11}{07}{46}{83} & \dec{-57}{07}{49}{1} & binary \\
GD1347.06.847 & \ra{11}{16}{01}{90} & \dec{-58}{29}{35}{7} & - \\
GD1348.11.709 & \ra{11}{19}{32}{94} & \dec{-59}{20}{59}{8} & - \\
GD1342.05.1150 & \ra{11}{26}{04}{85} & \dec{-60}{22}{11}{8} & - \\
GD1330.07.26061 & \ra{11}{43}{05}{94} & \dec{-60}{09}{13}{0} & incomplete light curve \\
GD1326.12.22665 & \ra{11}{52}{50}{92} & \dec{-62}{50}{56}{6} & - \\
GD1322.14.26962 & \ra{11}{54}{46}{63} & \dec{-66}{02}{32}{1} & - \\
\dots & \dots & \dots & \dots \\
\enddata
\tablecomments{Equatorial coordinates are given for the epoch J2000. (This table is available in its entirety in machine-readable form.)}
\end{deluxetable}

\clearpage

\section{Surface Density of Stars in OGLE GVS Fields}
\restartappendixnumbering 

\begin{deluxetable}{lrrrrrrrr}[h]
\tablecaption{Surface Density of Stars in OGLE GVS Fields \label{tab:stars}}
\tablehead{
\colhead{Field} & \colhead{R.A.} & \colhead{Decl.} & \colhead{$l$} &
\colhead{$b$} & \colhead{$\Sigma_{20}$} & \colhead{$\Sigma_{21}$} & \colhead{$N_{20}$} & \colhead{$N_{21}$} \\
\colhead{} & \colhead{(J2000, deg)} & \colhead{(J2000, deg)} & \colhead{(deg)} &
\colhead{(deg)} & \colhead{(arcmin$^{-2}$)} & \colhead{(arcmin$^{-2}$)} & \colhead{} & \colhead{} 
}
\startdata
DG1000.01 & 280.230 &  --3.893 &  28.269 &   0.560 &  118.3 &  219.5 &  19496 &  36173 \\
DG1000.02 & 280.076 &  --3.893 &  28.198 &   0.697 &  138.5 &  250.3 &  22829 &  41255 \\
DG1000.03 & 279.921 &  --3.893 &  28.127 &   0.834 &  131.5 &  252.9 &  21681 &  41699 \\
DG1000.04 & 279.767 &  --3.893 &  28.057 &   0.971 &  148.8 &  322.7 &  24524 &  53184 \\
DG1000.05 & 279.612 &  --3.893 &  27.986 &   1.108 &  153.1 &  343.9 &  25233 &  56687 \\
DG1000.06 & 279.458 &  --3.893 &  27.915 &   1.245 &  183.8 &  428.0 &  30288 &  70543 \\
DG1000.07 & 279.303 &  --3.893 &  27.845 &   1.382 &  163.4 &  344.1 &  26933 &  56735 \\
DG1000.08 & 280.385 &  --3.572 &  28.625 &   0.570 &  133.6 &  270.9 &  22024 &  44647 \\
DG1000.09 & 280.230 &  --3.572 &  28.554 &   0.707 &  168.2 &  354.1 &  27715 &  58331 \\
DG1000.10 & 280.076 &  --3.572 &  28.484 &   0.844 &  148.7 &  239.9 &  24507 &  39537 \\
DG1000.11 & 279.921 &  --3.572 &  28.413 &   0.981 &  154.0 &  254.7 &  25385 &  41976 \\
DG1000.12 & 279.767 &  --3.572 &  28.342 &   1.118 &  185.4 &  401.1 &  30551 &  66076 \\
DG1000.13 & 279.612 &  --3.572 &  28.272 &   1.255 &  221.0 &  456.7 &  36407 &  75243 \\
DG1000.14 & 279.458 &  --3.572 &  28.201 &   1.392 &  229.4 &  475.9 &  37780 &  78381 \\
DG1000.15 & 279.303 &  --3.572 &  28.130 &   1.529 &  206.0 &  451.1 &  33944 &  74339 \\
DG1000.16 & 279.149 &  --3.572 &  28.060 &   1.667 &  214.8 &  490.7 &  35390 &  80845 \\
DG1000.17 & 280.385 &  --3.273 &  28.891 &   0.707 &  101.6 &  200.8 &  16740 &  33069 \\
DG1000.18 & 280.230 &  --3.273 &  28.820 &   0.844 &  125.8 &  279.8 &  20719 &  46082 \\
DG1000.19 & 280.076 &  --3.273 &  28.750 &   0.981 &  149.4 &  253.6 &  24606 &  41775 \\
DG1000.20 & 279.921 &  --3.273 &  28.679 &   1.118 &  151.6 &  347.7 &  24976 &  57266 \\
\dots & \dots & \dots & \dots & \dots & \dots & \dots & \dots & \dots  \\
\enddata
\tablecomments{Here $\Sigma_{20}$ and $\Sigma_{21}$ are the surface densities of stars brighter than $I=20$ and 21, respectively, and $N_{20}$ and $N_{21}$ are the numbers of stars brighter than $I=18$ and 21, respectively. We note that the subfield (reference image) area may be slightly larger than the area covered by a single CCD detector because the reference image is the sum of a few frames that may be somewhat offset. (This table is available in its entirety in machine-readable form.)}
\end{deluxetable}

\end{document}